\documentclass[aps,prc,preprint,showpacs,showkeys,floatfix,nofootinbib,%
superscriptaddress]{revtex4}
\usepackage{graphicx}
\usepackage{multirow}
\usepackage{bm}

\def\be{\begin{equation}} \def\ee{\end{equation}} \def\bea{\begin{eqnarray}}
\def\eea{\end{eqnarray}} 

\newcommand{\eqn}[1]{\label{eq:#1}}
\newcommand{\refeq}[1]{(\ref{eq:#1})}

\newcommand{\Eq}{Eq.~\refeq}
\newcommand{\Eqs}[2]{Eqs.~(\ref{eq:#1}) and (\ref{eq:#2})}

\def\1s0{{}^1S_0}
\def\ts1{{}^3S_1}
\def\td1{{}^3D_1}
\def\ltap{\ \raise.3ex\hbox{$<$\kern-.75em\lower1ex\hbox{$\sim$}}\ }
\def\gtap{\ \raise.3ex\hbox{$>$\kern-.75em\lower1ex\hbox{$\sim$}}\ }
\def\ket#1{\vert#1\rangle}
\def\bra#1{\langle#1\vert}

\def\braketa#1{\langle#1\rangle}

\newcommand{\EG}{{\textrm{e.g.}}}

\date{\today}

\begin{document}

\title{Bridging over $p$-wave $\pi$-production and weak processes in
few-nucleon systems with chiral perturbation theory
}

\author{Satoshi X. Nakamura}\email{snakamura@triumf.ca}
\affiliation{Theory Group, TRIUMF,
4004 Wesbrook Mall, Vancouver, BC V6T 2A3, Canada}

\begin{abstract}
I study an aspect of chiral perturbation theory ($\chi$PT) which
 enables one to ``bridge'' different reactions.
That is, an operator fixed in one of the reactions can then be used to
 predict the other.
For this purpose,
I calculate the partial wave amplitude for the $p$-wave pion
 production ($pp\to pn\pi^+$)
using the pion production operator 
from the lowest and the next nonvanishing orders.
The operator includes
a contact operator whose coupling has been fixed
 using a matrix element of a low-energy weak
 process ($pp\to de^+\nu_e$).
I find that this operator does not
 reproduce the partial wave amplitude extracted from experimental data,
showing that the bridging over the reactions with significantly
 different kinematics is not necessarily successful.
I study the dependence of the amplitude on the various
 inputs such as the $NN$ potential, the $\pi N\Delta$ coupling, and the
 cutoff.
I argue the importance of a higher order calculation.
In order to gain an insight into a higher order calculation,
I add a higher order counter term to the operator used above, and fit the
 couplings to both the low-energy weak process and the pion production.
The energy dependence of the partial wave amplitude for the pion
 production is described by the operator consistently with the data.
However, I find a result which tells us to be careful about the
 convergence of the chiral expansion for the $pp\to pn\pi^+$ reaction.
\end{abstract}

\preprint{TRI-PP-07-25}

\maketitle

\section{Introduction}\label{sec_intro}

Applications of the chiral perturbation theory ($\chi$PT) to few-nucleon
systems have been of great interest in the last decade.
For reviews, see, {\it e.g.}, Refs.~\cite{review_eft}.
Particularly, a comprehensive review for the $\pi$ production
is given in Ref.~\cite{hanhart_review}, and for the electroweak
processes in Ref.~\cite{kubodera_review}.
A powerful aspect of $\chi$PT is that it provides us with a bridge between
different reactions in a model-independent way.
This means that coupling constants 
(the so-called low-energy constants, LECs)
fixed using experimental data
for one of the reactions
can then be used in the other.
An interesting interaction in this context is
\begin{eqnarray}
\eqn{d-term}
 {\cal L} = -2 d_1 N^\dagger S \cdot u N N^\dagger N \ ,
\end{eqnarray}
with
\begin{eqnarray}
 f_\pi u_\mu = -\tau_a \partial_\mu \pi_a 
- \epsilon_{3 ab} V_\mu \pi_a \tau_b 
+ f_\pi A_\mu + \cdots . 
\end{eqnarray}
The spin operator is $S$,
and the external vector (axial) current is $V_\mu$ ($A_\mu$).
The constant $d_1$ is a LEC.
This contact interaction between the nucleons
contributes to the three-nucleon force\cite{huber}, the $p$-wave
$\pi$-production ($pp\to pn\pi^+$\cite{hanhart_p-pi}, $pp\to d\pi^+$),
the radiative pion capture on the deuteron
($\pi^-d\to\gamma nn$)\cite{GP_PRL,GP}, 
and the weak processes
in few-nucleon systems,
such as tritium $\beta$-decay and the proton-proton fusion
($pp\to de^+\nu_e$)\cite{park}, the neutrino-induced disintegration of
the deuteron ($\nu_e d\to e^-pp$, $\nu d\to \nu pn$)\cite{ando1},
and the muon capture by the deuteron ($\mu^- d\to \nu_\mu nn$)\cite{ando2}.
The contact term in \Eq{d-term} makes the connection among these
reactions, which may be referred to as the two-body analog of the
Goldberger-Treiman relation,
as stated in Ref.~\cite{GP_PRL}.
If all couplings except for $d_1$ have been fixed using experimental data for
$\pi N$ and $NN$ elastic scattering, then
one can fix $d_1$ using one of the above reactions,
and predict the others.

There have been several such calculations which I will refer to as the
``bridging program''.
One of them was done by Park {\it et al.}~\cite{park}, where they
fixed $d_1$ using the experimental tritium $\beta$-decay rate, and
calculated, with no free parameters, the weak proton capture by a proton
(or $^3$He).
In this case, all the reactions are low-energy weak processes, and the
kinematics are relatively similar. Therefore, the bridging program
is expected to work well.
In another work due to Hanhart {\it et al.}~\cite{hanhart_p-pi},
the authors calculated the partial wave amplitude for the
$p$-wave $\pi$-production ($pp\to pn\pi^+$), and showed that
the use of $d_1$ fixed by three-nucleon observables~\cite{huber} 
also consistently reproduces the partial wave amplitude
extracted from experimental data~\cite{flammang}.
Although this result seems to be satisfactory, the bridging program
in this work was not done fully consistently. 
This is because they used a nuclear force~\cite{haidenbauer}
which is different from 
the nuclear force used for fixing $d_1$ in Ref.~\cite{huber}.
Because of the short-range nature of the $d_1$ term,
the $d_1$ value is largely dependent on the choice of the nuclear force.
Hanhart {\it et al.} also showed that the $p$-wave $\pi$-production
amplitude is rather largely dependent on the contribution from the $d_1$
term.
This indicates the importance of careful treatment of the $d_1$ term 
in the calculation; 
the nuclear force and the $d_1$ value have to be consistent.
Finally, I mention the bridging program done by G{\aa}rdestig {\it et
al.}~\cite{GP_PRL}, where they fixed the $d_1$ value using the matrix
element of $pp\to de^+\nu_e$ and then used it in calculating observables
for $\pi^- d\to \gamma nn$.
The two reactions are similar in kinematics of the $NN$ sector,
but are rather different in the momentum transfer, which might have to
be taken care of.
They are interested in extracting the neutron-neutron scattering length
from the $\pi^- d\to \gamma nn$ reaction and, for that purpose, they are
interested in the shape of the spectrum rather than the absolute value.
They found that the use of the $d_1$ term fixed in the above manner
significantly reduces the dependence of the shape on $R$ which is the
matching point between the long range one-pion-exchange potential and
the short range square well potential.
Although this result is a success of the bridging program, it is still
interesting to study the absolute value of the cross section in order to
more rigorously test the power of $\chi$PT.

In this work, I would like to more seriously investigate how reliably
the bridging program, an important aspect of $\chi$PT, works.
I believe that my investigation is important because there has been
sometimes an argument which supposes that the bridging program works,
even though it has not been quantitatively confirmed that
a bridging program over reactions with considerably different
kinematics works.
For this purpose, I calculate the partial wave amplitude for the
$p$-wave $\pi$-production in $NN$ collision ($pp\to pn\pi^+$), 
with $d_1$ fixed by an observable of a low-energy weak process.
This obviously provides a stringent test of $\chi$PT,
because the two reactions are strong and weak processes, and are 
quite different in kinematics.
My program is as follows.
I extend the operator of Ref.~\cite{park} by including the $\Delta$, 
which is known to play an important role in the $p$-wave $\pi$-production, 
and re-fit $d_1$ to an observable of a weak process.
Here, I fit $d_1$ to a matrix element of the axial vector current used
in Ref.~\cite{park}: the kinematics is that for the $pp \to de^+\nu_e$
reaction.
The $\pi$-production operators I use are the same as those used in
Ref.~\cite{hanhart_p-pi}, except that the $\Delta$ is treated differently.
In Ref.~\cite{hanhart_p-pi}, a coupled-channel equation was solved and
the $\Delta$ is included in the nuclear wave functions~\cite{haidenbauer}.
I use nuclear wave functions including only the nucleons as dynamical
degrees of freedom.
I take account of the $\Delta$ by including it in the $\pi$-production
operator.~\footnote{
Thus I treat the $\Delta$ perturbatively and do not fully take account of
some non-perturbative effect of the $\Delta$ in wave functions.
A calculation with a fuller account of the $\Delta$ might be
worthwhile doing to see a difference.
}
The $\pi$-production operator I use is based on a counting rule
proposed in Ref.~\cite{hanhart_p-pi,mod_counting}, 
which is different from Weinberg's counting\cite{Weinberg}, 
and the large initial on-shell momentum 
[$\sim \sqrt{m_N m_\pi}$, $m_N\ (m_\pi)$ : the nucleon (pion) mass]
is considered as a characteristic scale of the system.
I consider the operators up to next-to-leading order
[NLO, ${\cal O} (m_\pi/m_N)$].~\footnote{\label{foot_d1}
The operator used in Ref.~\cite{park} is based on Weinberg's counting
while the operator used in Ref.~\cite{hanhart_p-pi} is based on the counting
proposed in Ref.~\cite{hanhart_p-pi,mod_counting}.
Therefore, one may wonder whether the two operators contain different
mechanisms and thus the $d_1$ value is different in each case.
I will argue in Sec.~\ref{sec_di} that the $d_1$ value should be the
same in the both cases, up to the order I am working.
}

It is very interesting to see whether the operator
constructed using the low-energy weak process can reasonably describe
the $\pi$ production.
I will use several combinations of the inputs (the $\pi N\Delta$
coupling, the $NN$ potential, cutoff) in my calculation.
Even though
the $d_1$ value is fixed for each combination of the inputs
so that the low-energy weak process
is reproduced, one may expect a dependence of the $\pi$ production
amplitude on the combination because of 
the rather different kinematics between the two reactions.
I will study such a dependence.
In fact, we will see that this bridging program is not successful
when working with the NLO operators;
the partial wave amplitude of the $\pi$ production based on $\chi$PT
is not consistent with the data.
I will argue the importance of going to a higher order calculation.
In order to, even roughly, explore a result of a higher order
calculation, I add a higher order counter term and study a consequence.

This paper is organized as follows.
In Sec.~\ref{sec_op}, I present the chiral operator for the
$p$-wave $\pi$-production up to the order I work with.
I discuss the determination of the $d_1$ value using the
low-energy weak process in Sec.~\ref{sec_di}.
In Sec.~\ref{sec_cs}, I perform a multipole expansion of the operator,
and express the cross section in terms of the partial wave amplitudes.
Then, in Sec.~\ref{sec_result}, I present my result for the $p$-wave
$\pi$ production amplitudes and compare them with the data. The result
obtained with the higher order counter term is also presented.
Finally I summarize this work in Sec.~\ref{sec_summary}.

\section{chiral operator for $p$-wave $\pi$-production}\label{sec_op}
I present expressions of chiral operators which contribute to the
$p$-wave $\pi$-production.
I basically follow Ref.~\cite{hanhart_p-pi}, except for the treatment
of the $\Delta$ degree of freedom, and the way how the high momentum
components are cut off.
I start with the chiral interaction Lagrangian given in
Ref.~\cite{kolck_lagrangian}. 
By keeping terms relevant to my calculation, I have
\begin{eqnarray}
\eqn{l0}
 {\cal L}^{(0)}_{\rm int} & = & 
-\frac{1}{4 f_{\pi}^{2}}  N^{\dagger}
\bm{\tau} \cdot (\bm{\pi}\times\dot{\bm{\pi}})N +\frac{g_{A}}{2 f_{\pi}} 
         N^{\dagger}(\bm{\tau}\vec{\sigma}\cdot\vec{\nabla}\bm{\pi})N
                                               \nonumber \\
    &   & 
          +\frac{h_{A}}{2 f_{\pi}}N^{\dagger}(\bm{T}
          \vec{S}\cdot\vec{\nabla}\bm{\pi})\Delta  + {\rm h.c.}\ ,
\end{eqnarray} 
and
\begin{eqnarray}
\eqn{l1}
&& {\cal L}_{\rm int}^{(1)} 
        =\frac{i}{8m_{N}f_{\pi}^{2}} 
        N^{\dagger}\bm{\tau}\cdot
        (\bm{\pi}\times\vec{\nabla}\bm{\pi})\cdot\vec{\nabla}N 
         -\frac{c_3}{f_{\pi}^{2}}N^{\dagger}
        (\vec{\nabla}\bm{\pi})^{2}N                           \nonumber \\
  &   & -N^{\dagger}
   \frac{\bar c_4}{2f_{\pi}^{2}}
  \vec{\sigma}\cdot 
        \vec{\nabla}{\bm{\pi}} \times\vec{\nabla}
        \bm{\pi}\cdot\bm{\tau} N
        -\frac{ig_{A}}{4 m_{N} f_{\pi}}N^{\dagger}\bm{\tau}\cdot\dot{\bm{\pi}}
	\vec{\sigma}\cdot\vec{\nabla}N    \nonumber   \\  
  &   &        -\frac{i h_{A}}{
        4 m_{N} f_{\pi}}
        N^{\dagger}\bm{T}\cdot\dot{\bm{\pi}}\vec{S}\cdot\vec{\nabla}
        \Delta                      
        -\frac{d_1}{f_{\pi}} 
        N^{\dagger}\bm{\tau}\vec{\sigma}\cdot\vec{\nabla}\bm{\pi} N\,
        N^{\dagger}N \nonumber \\
&   &        -\frac{d_2}{2 f_{\pi}}  
        \vec{\nabla}\bm{\pi} \times  
        N^{\dagger}\vec{\sigma}\bm{\tau} N\;\cdot
        N^{\dagger}\vec{\sigma}\bm{\tau} N 
        + {\rm h.c.},
\end{eqnarray}
with $\bar c_4 = c_4+\frac{1}{4m_N}$.
The nucleon's spin (isospin) operator is $\vec{\sigma}$ ($\bm{\tau}$), while
the $N\Delta$ transition spin (isospin) operator is $\vec{S}$ ($\bm{T}$).
I use the pion decay constant $f_\pi =$ 93~MeV and the nucleon mass,
$m_N =$ 939~MeV.
Regarding the parameters ($g_A, h_A, c_3, c_4$), I follow
Ref.~\cite{krebs}, where Krebs {\it et al.} constructed a chiral nuclear force
including the $\Delta$ explicitly.
The axial coupling constant is $g_A$ (= 1.27).
For the $\pi N\Delta$ coupling, I use two choices: $h_A = 2.10$ from
the $\Delta$-decay width\cite{fettes}, 
and $h_A = 3 g_A/\sqrt{2} = 2.68$ from large
$N_C$.~\footnote{The definition of the $\pi N\Delta$ coupling constant
here is different from that of Ref.~\cite{krebs} by a factor of 2:
$h_A$~(this~work) $= 2 h_A$~(Ref.~\cite{krebs}).
}
In Ref.~\cite{krebs}, the authors calculated the $s$- and $p$-wave 
$\pi N$ scattering threshold parameters at next-to-leading order with taking
account of the $\Delta$, and fit the
couplings ($c_3$, $c_4$ and others) to the analysis of
Matsinos\cite{em98}.
The result is
$c_3 = -1.87$ GeV$^{-1}$, $c_4 = 1.87$ GeV$^{-1}$ for $h_A = 2.10$, and 
$c_3 = -0.79$ GeV$^{-1}$, $c_4 = 1.33$ GeV$^{-1}$ for $h_A = 2.68$.
The remaining unknown LECs are $d_1$ and $d_2$ which will be determined
in the next section.

\begin{figure}[t]
\includegraphics[width=6in]{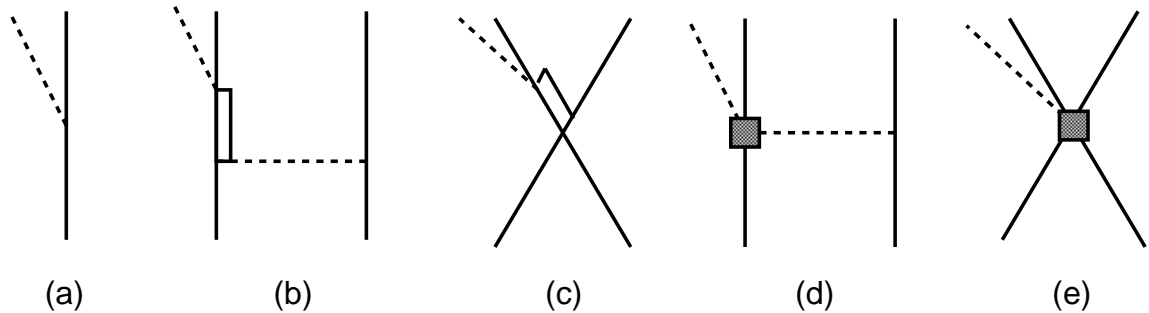}
\caption{\label{fig_diag}
The $p$-wave $\pi$-production operators up to NLO in $\chi$PT.
Dashed lines denote pions and the solid lines nucleons, the double lines
 $\Delta$. Vertices without (with) the shaded box arise from the leading
 (subleading) order Lagrangian.
}
 \end{figure}
I use the following $\pi$-production operator derived from the above
interaction Lagrangian.
The leading order [LO, ${\cal O}(1)$] operator is the one-body direct production of the pion
with the isospin state $a$ off the nucleon [Fig.~\ref{fig_diag}(a)], 
\begin{eqnarray}
\eqn{1b}
 O_{\rm 1B} &=& i \frac{g_A}{2f_\pi} (2\pi)^3
  \delta^{(3)}(\bm{p}_2^\prime - \bm{p}_2) \tau_1^a
  \bm{\sigma}_1\cdot\bm{q}_\pi\  +\ (1 \leftrightarrow 2) \ ,
\end{eqnarray}
where $\bm{q}_\pi$ is the momentum of the emitted pion.
The quantity $\bm{p}_i$ ($\bm{p}_i^\prime$) is the momentum of the incoming
(outgoing) $i$-th nucleon.
Another LO mechanism I consider is the $\Delta$-excitation followed
by the $\pi$ emission.
In Ref.~\cite{hanhart_p-pi}, the authors used the wave function which
explicitly includes the $\Delta$ component, and considered the one-body
operator which produces the pion with the $\Delta$ deexcited to the
nucleon.
Because I use nuclear wave functions with only the nucleonic degrees of
freedom, I alternatively use a two-body operator in which the $\Delta$
is excited either by the $\pi$-exchange [Fig.~\ref{fig_diag}(b)], or by
a contact interaction between the nucleons [Fig.~\ref{fig_diag}(c)].
The former is given by
\begin{eqnarray}
\eqn{pi-d}
 O_{\Delta\pi} &=& \frac{i}{36}\ 
 \frac{g_A h_A^2}{f_\pi^3}\ 
 \frac{\bm{\sigma}_2\cdot\bm{k}}{m_\pi^{\prime 2} + k^2}\ 
\frac{\left( 4\tau_2^a\bm{k} 
- (\bm{\tau}_1\times\bm{\tau}_2)^a \bm{\sigma}_1 \times \bm{k}\ 
\right)\cdot \bm{q}_\pi}{m_\Delta - m_N - p_o^2/m_N +
  (\bm{p}^\prime+\bm{q}_\pi/2)^2/2\mu}\  +\ (1 \leftrightarrow 2) \ ,
\end{eqnarray}
where $m_\Delta$ ($= 1232$~MeV) is the mass of the $\Delta$, 
and $p_o$ is the initial on-shell relative $NN$ momentum.
The quantities, $\mu$, $k$, $p^\prime$ are respectively defined by
$\mu\equiv m_Nm_\Delta/(m_N+m_\Delta)$,
$\bm{k}\equiv \bm{p}_2-\bm{p}_2^\prime$
and $\bm{p}^\prime\equiv (\bm{p}^\prime_1-\bm{p}_2^\prime)/2$.
I assume the equal energy sharing between the nucleons and
$m_\pi^{\prime 2}\equiv (3m_\pi^2-q_\pi^2)/4$, $m_\pi$ = 138~MeV.
In Appendix,
I explain how I treat the energy denominator in \Eq{pi-d} in my calculation.
An expression for the diagram in Fig.~\ref{fig_diag}(c) is
\begin{eqnarray}
\eqn{ct-d}
 O_{\Delta{\rm CT}} &=& - \frac{i}{9}\ 
 \frac{h_A C_2^{N\Delta}}{f_\pi}\ 
\frac{\left( 
4 \, \tau_2^a \,
         ( {\bm \sigma}_2\!\times\!\bm{k}) 
         \!\times\!\bm{k}_j
     -( {\bm \tau}_1 \times {\bm \tau}_2)^a \,
      {\bm \sigma}_1\!\times\!
    \left [ ( {\bm \sigma}_2\!\times\!\bm{k}) 
      \!\times \bm{k} \right ]
\right)\cdot \bm{q}_\pi} 
{m_\Delta - m_N - p_o^2/m_N +
  (\bm{p}^\prime+\bm{q}_\pi/2)^2/2\mu} \nonumber\\
  &+&  (1 \leftrightarrow 2) \ ,
\end{eqnarray}
where I have used the contact $NN\to N\Delta$
interaction with two derivatives,
\begin{eqnarray}
\eqn{n-delta}
{\cal L}_{NN\to N\Delta}\! &=&\!
- i C_2^{N\Delta} \Delta^\dagger \bm{T} \left(
\vec{S}\cdot(\stackrel{\rightarrow}{\nabla} + \stackrel{\leftarrow}{\nabla})
\vec{\sigma}
-\vec{S} \vec{\sigma}\cdot
(\stackrel{\rightarrow}{\nabla} + \stackrel{\leftarrow}{\nabla})
\right)N\! \cdot\!
N^{\dagger}\left(\vec{\sigma}\times
(\stackrel{\rightarrow}{\nabla} + \stackrel{\leftarrow}{\nabla})
\right)\bm{\tau}N \ .
\end{eqnarray}
It is noted that a contact $NN\to N\Delta$
interaction without derivative does not contribute to the transition
under consideration. 
I use the $C_2^{N\Delta}$ value taken from the resonance saturation of
the $\rho$-exchange $NN\to N\Delta$ potential used in several
phenomenological models\cite{schiavilla}:
\begin{eqnarray}
\eqn{c2nd}
h_A C_2^{N\Delta} &=& g_A \frac{18}{25}\ \frac{g_{\rho NN}^2}{m_N^2}\ 
(1 + \kappa_\rho)^2\ \frac{1}{m_\rho^2}\ ,
\end{eqnarray}
where the $\rho NN$ vector coupling is $g_{\rho NN}$
($g_{\rho NN}^2/4\pi = 0.5$)
and the tensor coupling is $\kappa_\rho$ ($=6.6$). 
The factor at the end, 
$1/m_\rho^2$ ($m_\rho$ = 770~MeV), is the static limit
of the $\rho$-meson propagator.
Although Eq.~(\ref{eq:n-delta}) is not a general 
contact $NN\to N\Delta$ interaction with two derivatives,
one may take account of the most important part of the $NN\to N\Delta$
contact interaction by invoking a meson-exchange model, and the
remaining part may be phenomenologically absorbed by the contact
$d_i$ ($i=1,2$) terms.
Obviously, my treatment of the $\Delta$ is rather phenomenological.
However, a construction of a nuclear force from a chiral
Lagrangian with the $\pi$, $N$ and $\Delta$, which is yet to be done
\footnote{Such a nuclear force for the peripheral wave has been
constructed recently~\cite{krebs}.},
is necessary to determine the $C_2^{N\Delta}$ value.
I believe that my treatment is one of what one can do best
for the moment, and expect a fully consistent calculation in future.

Next I discuss next-to-leading order [NLO, ${\cal O}(m_\pi/m_N)$] terms which consist of
four types.
One of them is the recoil correction to the LO terms.
The recoil corrections to the one-body term [\Eq{1b}],
$\pi\Delta$ term [\Eq{pi-d}],
and contact-$\Delta$ term [\Eq{ct-d}] are respectively given by
\begin{eqnarray}
\eqn{recoil-1b}
 O_{\rm 1B, recoil} &=& - i \frac{g_A \omega_\pi}{4 m_N f_\pi} (2\pi)^3
  \delta^{(3)}(\bm{p}_2^\prime - \bm{p}_2) \tau_1^a
  \bm{\sigma}_1\cdot (\bm{p}_1 + \bm{p}_1^\prime)\  +\ (1 \leftrightarrow 2) \ ,
\end{eqnarray}
\begin{eqnarray}
\eqn{recoil-pi-d}
 O_{\Delta\pi, {\rm recoil}} &=& -\  \frac{i}{72}\ 
\frac{\omega_\pi}{m_N}
 \frac{g_A h_A^2}{f_\pi^3}\ 
 \frac{\bm{\sigma}_2\cdot\bm{k}}{m_\pi^{\prime 2} + k^2}
\nonumber\\
&\times&
\frac{ 4 \tau_2^a \bm{k}\cdot (\bm{p}_1+\bm{p}_1^\prime) 
- (\bm{\tau}_1\times\bm{\tau}_2)^a \left[
 (\bm{\sigma}_1 \times \bm{k})\cdot (\bm{p}_1+\bm{p}_1^\prime) 
+ 2 i k^2 \right]
}{m_\Delta - m_N - p_o^2/m_N +
  (\bm{p}^\prime-\bm{q}_\pi)^2/2\mu}\\\nonumber
  &+& (1 \leftrightarrow 2) \ ,
\end{eqnarray}
\begin{eqnarray}
\eqn{recoil-ct-d}
 O_{\Delta{\rm CT,recoil}} &=&  \frac{i}{18}\ 
 \frac{\omega_\pi}{m_N}\
 \frac{h_A C_2^{N\Delta}}{f_\pi}
\frac{1}{m_\Delta - m_N - p_o^2/m_N +
  (\bm{p}^\prime-\bm{q}_\pi)^2/2\mu}\nonumber\\
&\times& \left\{  \tau_2^a \,\left[ - 2 k^2 \bm{k}\cdot (\bm{\sigma}_1\times\bm{\sigma}_2)
         + 4(( {\bm \sigma}_2\!\times\!\bm{k}) 
         \!\times\!\bm{k})\cdot (\bm{p}_1+\bm{p}_1^\prime) \right]\right.\nonumber\\
     &-& \left.( {\bm \tau}_1 \times {\bm \tau}_2)^a \,
      \left(\bm{\sigma}_1\!\times\!
    \left( ( {\bm \sigma}_2\!\times\!\bm{k}) 
      \!\times \bm{k} \right)\right)\cdot (\bm{p}_1+\bm{p}_1^\prime) 
\right\}
+\ (1 \leftrightarrow 2) \ ,
\end{eqnarray}
where $\omega_\pi$ is the energy of the emitted pion.
The second type of the NLO operator is a pion rescattering through
either the $c_3$ term, or the $c_4$ term or
the Galilean correction to the Weinberg-Tomozawa term [the first
term in \Eq{l1}]. They are graphically represented by
Fig.~\ref{fig_diag}(d), and their expressions are given by
\begin{eqnarray}
\eqn{c3}
 O_{c_3} &=& -i \frac{c_3 g_A}{f_\pi^3}
 \frac{\bm{\sigma}_2\cdot\bm{k}}{m_\pi^{\prime 2} + k^2}
\bm{k} \cdot \bm{q}_\pi\tau_2^a \  +\ (1 \leftrightarrow 2) \ , 
\end{eqnarray}
\begin{eqnarray}
\eqn{c4}
 O_{c_4} &=& -i \frac{ \bar{c}_4 g_A}{2f_\pi^3} 
 \frac{\bm{\sigma}_2\cdot\bm{k}}{m_\pi^{\prime 2} + k^2}
(\bm{\sigma}_1 \times \bm{k}) \cdot \bm{q}_\pi
(\bm{\tau}_1\times\bm{\tau}_2)^a \  +\ (1 \leftrightarrow 2) \ ,
\end{eqnarray}
\begin{eqnarray}
\eqn{galilean-wt}
 O_{{\rm WT, Galilean}} &=& \frac{g_A}{16 m_N f_\pi^3}
 \frac{\bm{\sigma}_2\cdot\bm{k}}{m_\pi^{\prime 2} + k^2}
(\bm{q}_\pi-\bm{k})\cdot (\bm{p}_1+\bm{p}_1^\prime)
({\bm \tau}_1 \times {\bm \tau}_2)^a\  +\ (1 \leftrightarrow 2) \ . 
\end{eqnarray}
For a convenience, I decompose
the last term as
$O_{{\rm WT, Galilean}} = O_{{\rm WT1(G)}} + O_{{\rm WT2(G)}}$ with
\footnote{
When convoluted with the wave functions,
$O_{\rm WT, Galilean}$ gives a non-vanishing amplitude
in the soft pion limit.
(The tree amplitude vanishes in the soft pion limit.)
This is not consistent with chiral symmetry.
This problem is similar to that found in Ref.~\cite{gardestig} in the
context of the $s$-wave pion production in the $NN$ collision.
The solution to this problem was given in Ref.~\cite{lensky}.
Probably, the problem here is also resolved by the same mechanism found
in Ref.~\cite{lensky}; \EG, one loop diagram formed by 
$O_{\rm WT, Galilean}$
and the one-pion-exchange potential is (partly)
canceled out by other irreducible pion loop diagrams, leaving a contribution
consistent with the chiral symmetry.
One may take some prescription to maintain the chiral symmetry.
However, I use \Eq{galilean-wt} without modification because this term
gives a small contribution (a few percents) to an amplitude for the
$pp\to pn\pi^+$ reaction; the modification will not significantly change
results.
}
\begin{eqnarray}
\eqn{galilean-wt2}
O_{{\rm WT1(G)}} &=& \frac{g_A}{16 m_N f_\pi^3}
 \frac{\bm{\sigma}_2\cdot\bm{k}}{m_\pi^{\prime 2} + k^2}
(2\bm{p}^\prime + \bm{k})\cdot \bm{q}_\pi
({\bm \tau}_1 \times {\bm \tau}_2)^a\
+\ (1 \leftrightarrow 2) \ , \\
O_{{\rm WT2(G)}} &=& \frac{g_A}{16 m_N f_\pi^3}
 \frac{\bm{\sigma}_2\cdot\bm{k}}{m_\pi^{\prime 2} + k^2}
(- k^2 -2 \bm{p}^\prime\cdot\bm{k})
({\bm \tau}_1 \times {\bm \tau}_2)^a\
+\ (1 \leftrightarrow 2) \ .
\end{eqnarray}
The third type of the NLO terms is a pion rescattering via the
Weinberg-Tomozawa term:
\begin{eqnarray}
\eqn{wt}
 O_{{\rm WT}} &=& - \frac{3 g_A\omega_\pi}{16 f_\pi^3}
 \frac{\bm{\sigma}_2\cdot\bm{k}}{m_\pi^{\prime 2} + k^2}
({\bm \tau}_1 \times {\bm \tau}_2)^a\  +\ (1 \leftrightarrow 2) \ .
\end{eqnarray}
Finally, the fourth type is a pion emission from the contact terms
[the $d_1$ and $d_2$ terms in \Eq{l1}, Fig.~\ref{fig_diag}(e)]:
\begin{eqnarray}
\eqn{d_i}
 O_{d} &=& -i \left(
\frac{d_1}{f_\pi} \bm{\sigma}_1\tau_1^a
+ \frac{d_2}{2f_\pi} (\bm{\sigma}_1\times\bm{\sigma}_2)
({\bm \tau}_1 \times {\bm \tau}_2)^a
\right)\cdot \bm{q}_\pi
 +\ (1 \leftrightarrow 2) \ .
\end{eqnarray}

Starting with the operators presented above, I take the following
procedure
to obtain the momentum space ($p$-space) operator contributing to
the $^1\!S_0\to ^3\!S_1$-$^3\!D_1$ transition.
At first, I perform Fourier transformation of the above operators to
represent them in the coordinate space ($r$-space). 
In the Fourier transformation,
I multiply a Gaussian cutoff function, 
$\exp\left(-k^2/\Lambda_G^2\right)$, to the operators other than the
one-body operators; I use $\Lambda_G = $ 2~GeV.
In the $r$-space, I perform the multipole expansion of the
operators, and evaluate the matrix elements for the spin, isospin and
angular parts of the operators.
The remaining radial part of the operators is converted to the $p$-space using
Fourier transformation, and I obtain the radial part of the $p$-space
operator, $O(p',p)$, where $p$ ($p'$) is off-shell relative momentum of
the incoming (outgoing) two nucleons.
I classify the operators into two groups:
\begin{center}
 \begin{tabular}{lcl}%
 Group I &:& $O_{\Delta\pi}$, $O_{\Delta{\rm CT}}$, $O_{c_3}$
 , $O_{c_4}$, $O_{d}$\\
 Group II &:& $O_{\rm 1B}$, $O_{\rm 1B,recoil}$, 
 $O_{\Delta\pi,{\rm  recoil}}$, $O_{\Delta{\rm CT,recoil}}$
 , $O_{\rm WT, Galilean}$, $O_{\rm WT}$\\
 \end{tabular}
\end{center}
I introduce a sharp cutoff $\Lambda$ for the operators belonging to
Group I such that
\begin{eqnarray}
 O_\Lambda (p',p) &=&  O(p',p)\ ,  \qquad {\rm for}\ \ p\le\Lambda\ \ {\rm and}\
  p'\le\Lambda\ , \nonumber\\
 O_\Lambda(p',p) &=&  0\ ,  \qquad\qquad\qquad {\rm otherwise}\ .
\end{eqnarray}
I do not apply the sharp cutoff to the operators of Group II.
As we will see in the next section where the LECs $d_i$ are determined
using low-energy weak processes,
the renormalization of $d_i$
only takes care of high momentum components (larger than $\Lambda$) of the
operators in Group I.
The LECs $d_i$ contain the same physics for both the reactions bridged.
Therefore, I retain the high momentum components of the
operators in Group II in my calculation.~\footnote{
One may also apply the sharp cutoff to the Group II operators,
which amounts to cutting off a higher order effect.
}
As representatives, I choose $\Lambda =$ 500, 600 and 800~MeV.
I use the $p$-space operator explained above because of its usefulness for
the renormalization group (RG) analysis which will be done later.

For the purpose of the multipole expansion of the $\pi$ production
operator, which will be done in Sec.~\ref{sec_cs}, I express the
above $\pi$ production operators as follows:
\begin{eqnarray}
\eqn{ox}
 O_X &=& - \vec{O}_X\cdot \vec{Q}, \qquad X = {\rm 1B},\ \Delta\pi,\
  \Delta{\rm CT},\ c_3,\ c_4,\ {\rm WT1(G)},\ d \\
\eqn{oy}
 O_Y &=& O_Y^0\ \; Q^0, \qquad Y = {\rm recoil\ terms},\ {\rm WT2(G)},\ {\rm WT}
\end{eqnarray}
where $\vec{Q}=i\vec{q}_\pi/2f_\pi$ and $Q^0 = i\omega_\pi/2 f_\pi$.
I will refer to $\vec{O}_X$ as the spatial component of the $\pi$
production operator while $O_Y^0$ as the time component.

\section{Determination of $d_i$}\label{sec_di}

In this section, I determine the LECs $d_i$ in \Eq{d_i} using an
observable of a low-energy weak process.
The $d_i$ terms contribute to
the $^1S_0\to {}^3\!S_1$ transition and
appear as a single linear combination, $\tilde{d}\equiv d_1+2d_2$.
Because $d_1$ and $d_2$ are not separately determined 
by considering the $^1S_0\to ^3\!S_1$ transition only,
I determine $\tilde{d}$ rather than $d_1$ and $d_2$ individually.

In Ref.~\cite{park}, the authors fixed $\tilde{d}$
using the experimental data of the tritium $\beta$-decay rate.~\footnote{
The authors fixed dimensionless constant $\hat{d}_R$ rather than
$\tilde{d}$. The two quantities are related through
$\hat{d}_R = (m_N f_\pi^2/g_A) \tilde{d} + \hat{c}_3/3 + 2\hat{c}_4/3 + 1/6$, 
with $c_{3,4} = \hat{c}_{3,4}/m_N$.
}
They derived the axial vector current contributing the tritium
$\beta$-decay from the chiral Lagrangian.
They did not explicitly consider the $\Delta$ as a degree of freedom.
The spatial component of the axial vector currents used in
Ref.~\cite{park} are obtained from the one-body operator [\Eq{1b}]
and Group I operators without the $\Delta$
[Eqs.~(\ref{eq:c3}), (\ref{eq:c4}) and (\ref{eq:d_i})], with
the factor of $(i\bm{q}_\pi/2f_\pi)$ being eliminated and
$m_\pi^\prime$ replaced by $m_\pi$.
Accordingly,
the parameters ($c_i$) used in Ref.~\cite{park} are different from those
presented in the previous section.
Although they additionally included some other operators which
give a negligible contribution, I do not 
consider them in the following.
They multiply a Gaussian cutoff function,
$\exp\left(-k^2/\Lambda_\chi^2\right)$, to the operators,
with $\Lambda_\chi =$ 500, 600 and 800~MeV. 
For each choice of $\Lambda_\chi$, they fixed $\tilde{d}$ so
that the tritium $\beta$-decay rate is reproduced.
They used the AV18 $NN$ potential~\cite{av18} supplemented by the
Urbana-IX three-nucleon potential~\cite{urbana} when calculating the
matrix element for the tritium $\beta$-decay.

Here, I need to re-fit the value of $\tilde{d}$ in ``my'' axial
current operator.
My operator is
the axial current operator used in Ref.~\cite{park}
plus the $\Delta$-excitation current.
The $\Delta$-excitation current is obtained from \Eqs{pi-d}{ct-d}
by eliminating the factor of $(i\bm{q}_\pi/2f_\pi)$ and
$m_\pi^\prime \to m_\pi$.
The way of cutting off the high momentum states is also different from
that used in Ref.~\cite{park}, as has been 
discussed in the previous section.
I do not directly use the tritium $\beta$-decay rate but take an
indirect way to fix $\tilde{d}$, as discussed in the next paragraph.

I start with a benchmark calculation to which the $\tilde{d}$ value in
my operator is fitted.
For that purpose,
I use the same spatial axial current operator used in Ref.~\cite{park}
(the same couplings and the same cutoff but without
the negligible operators) to calculate a matrix element for the
$^1S_0\to ^3\!S_1$-$^3\!D_1$ transition in two-nucleon system. 
This matrix element is the benchmark to which $\tilde{d}$ is
re-fitted so that ``my'' operator reproduces the same matrix element.
In the benchmark calculation,
I use the AV18 potential for consistency, and 
choose a kinematics with $T_{NN}^i =$ 0.5~MeV 
($T_{NN}^i$ : the initial on-shell kinetic energy of the relative
motion), the deuteron final state, and $q =$ 2.5~MeV
($q$ : the momentum transfer from
the two-nucleon system to the external current).
This kinematics is for the low-energy $pp\to de^+\nu_e$ reaction.
I use the proton-neutron interaction to generate the initial wave
functions, and therefore do not consider
the Coulomb interaction and the isospin violation effect.
I calculate the matrix element with different choices of the Gaussian
cutoff, $\Lambda_\chi =$ 500, 600 and 800~MeV.
I average the three matrix elements which have a slight cutoff dependence,
and regard the average as the benchmark.
When I calculate the matrix element of my operator,
I use several combinations of
the $NN$ potential, the $\pi N\Delta$ coupling ($h_A$) and
the corresponding $c_i$ values, and the sharp cutoff value ($\Lambda$).
I use the following $NN$ potentials: the CD-Bonn\cite{cdbonn}, the
AV18\cite{av18}, the Reid93\cite{nij}, the Nijmegen I\cite{nij} and the
chiral N$^3$LO\cite{n3lo} potentials.~\footnote{
I take the hybrid approach where a matrix element of the chiral
operator is sandwiched by wave functions obtained with a
phenomenological nuclear force. This approach is also referred to as
more effective effective field theory (MEEFT). An argument for employing 
MEEFT is given in Ref.~\cite{meeft}.
}
For each combination of these inputs,
the $\tilde{d}$ value is fitted to the benchmark,
and the result is given in
Table~\ref{tab_d}. In the table, I show the dimensionless coupling
$\hat{d} (\equiv m_N f_\pi^2\;  \tilde{d})$. 
\begin{table}[t]
\renewcommand{\arraystretch}{1.2}
\tabcolsep=2.3mm
\caption{\label{tab_d}
Dimensionless contact coupling, $\hat{d} (\equiv m_N f_\pi^2\; \tilde{d})$. 
The first column is the sharp cutoff value. 
The first row specifies the $NN$ potentials used.
For each $NN$ potential, the
 left side is $\hat{d}$ for $h_A =$ 2.10, while the right side is for
$h_A =$ 2.68.
}
 \begin{tabular}[t]{cccccccccccc}\hline
 $\Lambda$ &\multicolumn{2}{c}{CD-Bonn}
&\multicolumn{2}{c}{AV18}&\multicolumn{2}{c}{Reid93}
 &\multicolumn{2}{c}{Nij I}&\multicolumn{2}{c}{N$^3$LO}\\\hline
500& 0.18& 0.06&$-$0.01&$-$0.17&$-$0.48&$-$0.64& 0.29& 0.15&$-$0.03&$-$0.20\\	   
600& 0.62& 0.42& 0.77& 0.47& 0.17&$-$0.13& 0.93& 0.67& 0.60& 0.32\\	   
800& 1.74& 1.32& 4.36& 3.32& 3.16& 2.11& 2.75& 2.13& 1.18& 0.81\\\hline
 \end{tabular}
\end{table}
Although the $\tilde{d}$ value is adjusted so that the benchmark
is reproduced irrespective of changing 
the inputs, it would be expected that the
$\pi$-production amplitude, evaluated in a significantly different kinematics,
have the dependence on them.
I will study the dependence later.

Before closing this section, I discuss the issue mentioned in
footnote~\ref{foot_d1}.
In order for the bridging program to be meaningful, the $\tilde{d}$
term has to (implicitly) include the same mechanisms
(up to the external probe) for both of the reactions bridged.
If the $\tilde{d}$ term for one of the reactions includes a mechanism
which is explicitly considered in the other, 
the $\tilde{d}$ value should be different in each case.
In the case under consideration,
the axial current operator and the pion production operator are based on
different counting schemes, and
therefore there is a concern that
the two operators explicitly contain different
mechanisms.
In the following,
I will argue that the $\tilde{d}$ term includes the same physics for
both of the operators, up to the order I am working.
As we have seen in the above, all of the mechanisms for the axial
vector current are included in the pion production operator
(up to the external probe).
However,
the pion production operator contains some mechanisms which are
not included in the axial vector current; the recoil corrections,
WT and WT(G).
Among them, WT and WT(G) do not contribute to the weak process, and thus
the $\tilde{d}$ term fixed in the weak process does not contain these
mechanisms.
The recoil correction of the one-body axial current
[the counterpart of \Eq{recoil-1b}] has been considered in Ref.~\cite{park},
which means that
the $\tilde{d}$ term fixed in Ref.~\cite{park} does not contain this
mechanism.
Although I did not consider this mechanism
when fixing $\tilde{d}$,
I am safe because this mechanism gives only a negligible
contribution to the matrix element considered.
The recoil corrections of the $\Delta$ axial current
[the counterpart of \Eqs{recoil-pi-d}{recoil-ct-d}]
have not been considered in Ref.~\cite{park}.
However, the $\tilde{d}$ value does not depend on the inclusion of these
mechanisms because they give a negligible contribution to
the matrix element.
(In fact, these recoil corrections should be captured by another higher
order counter term.)
Therefore, up to the order I am working, the $\tilde{d}$ term contains
the same physics for both of the reactions bridged, and I can use the
$\tilde{d}$ term fixed in this section for calculating the pion
production amplitude.

\section{partial wave amplitude and cross section}\label{sec_cs}

In this section, I will express the cross section for
the $pp\to pn\pi^+$ reaction in terms of partial wave amplitudes.
\footnote{
I will work with the center-of-mass system of the initial $pp$ system
throughout this work.
}
For this purpose, I perform the standard multipole expansion of
the $\pi$ production operator\cite{multipole}, and a partial wave
expansion of the
initial and final $NN$ scattering wave functions.
Then, I use the partial wave amplitudes to
express the transition amplitude, and subsequently the cross section.
I perform these expansions in the $r$-space, which will be followed by
the conversion into the $p$-space.

\subsection{Partial wave amplitude}\label{sec_partial_amp}
The multipole operator for the time component of 
a $\pi$ production operator is defined by
\begin{eqnarray}
\eqn{op-charge}
T_C^{JM}({\cal O}) & = & \int d\bm{x} j_J(qx)Y_{JM}(\hat{\bm{x}})
{\cal O}^0(\bm{x}),
\end{eqnarray}
where $\bm{q} \equiv -\bm{q}_\pi$ ($q = |\bm{q}|$),
$j_J(qx)$ is the spherical Bessel function of order $J$,
and $\hat{\bm{x}}\equiv \bm{x}/|\bm{x}|$.
An $r$-space operator is ${\cal O}^0(\bm{x})$, the dependence on the
center-of-mass coordinate being eliminated; {\it e.g.},
${\cal O}^0(\bm{x}) = O^0(\bm{x})|_{\bm{R}=0}$.
For the spatial component of the $\pi$ production operator,
I show only the longitudinal multipole operator
because the electric and magnetic multipole operators give vanishing
contribution for the case in question:
\begin{eqnarray}
\eqn{op-longi}
  T_L^{JM}({\cal O}) & = & 
  \frac{i}{q} \int d\bm{x} 
  \bm{\nabla} [j_J(q x)Y_{JM}(\hat{\bm{x}})]
   \cdot \vec{{\cal O}}(\bm{x}). 
\end{eqnarray}
The transition amplitude for the $pp\to pn\pi^+$ reaction 
is written by
\begin{eqnarray}
\eqn{amp}
  T_{fi} = \bra{\psi_{f}}\!
\int d\bm{x}\    e^{i\bm{q}\cdot\bm{x}}
   \left[\sum_Y{\cal O}^0_Y(\bm{x}) Q^0 - 
   \sum_X\vec{{\cal O}}_X(\bm{x})\cdot\vec{Q}\right]
\ket{\psi_i}\ ,
\end{eqnarray}
where $\psi_i$ ($\psi_f$) is the nuclear wave function for the initial
(final) state, whose $r$-space representation will be given later.
In the summations in \Eq{amp}, $X$ ($Y$) takes various components of the
$\pi$ production operator specified in \Eq{ox} [\Eq{oy}].
The four-vector $Q$ has been defined in \Eqs{ox}{oy}.
Using the multipole operators presented above,
I can rewrite $T_{fi}$ as
\begin{eqnarray}
  T_{fi} &=&
\sum_{J_o M_o} 4 \pi i^{J_o}
  (-1)^{M_o}\bra{\psi_f}\left[ 
T_C^{J_oM_o} Q_C^{J_o-M_o}
+ T_L^{J_oM_o} Q_L^{J_o-M_o}
\right]\,\ket{\psi_i}\ ,
\label{eq_me-jl}
\end{eqnarray}
with
\begin{eqnarray}
  Q_C^{JM} & = & 
Y_{JM}(\hat{\bm{q}}) \  Q^0,\\
  Q_L^{JM} & = & 
\left( \sqrt{\frac{J}{2J+1}}\bm{Y}_{J-1 J
  M}(\hat{\bm{q}})- 
\sqrt{\frac{J+1}{2J+1}}\bm{Y}_{J+1 J
  M}(\hat{\bm{q}}) \right)\cdot \bm{Q} \ ,
\end{eqnarray}
and $\bm{Y}_{JLM}(\hat{\bm{q}})$ 
are the vector spherical harmonics.

Now I proceed to the partial wave expansion of the $NN$ wave function.
An $NN$ scattering wave function with the relative momentum $\bm{p}$, 
the third component of the spin
(isospin) of $i$-th nucleon being $s_i$ ($\tau_i$) is expanded as follows:
\begin{eqnarray}
\eqn{NNwave}
\psi(\bm{r})\!\! & = &\!\!
\sum_{\alpha,m}
 4\pi(1/2,s_{1},1/2,s_{2}|S \mu)
(1/2,\tau_{1},1/2,\tau_{2}|T,T^3)
(L m S \mu |J M) 
i^{L} Y_{L,m}^*(\hat{\bm{p}})
 \psi_{\alpha}(\bm{r}) ,
\end{eqnarray}
where the index $\alpha$  collectively denotes the quantum
numbers of a partial wave; 
$\alpha = \{J, L, S, T \}$ where 
$J, L, S$ are the total, orbital, total spin angular momenta of
the $NN$ system, respectively and $T$ is the total isospin.
The partial wave function ($\psi_\alpha$) is expressed as
\begin{eqnarray}
\eqn{partial_wave}
 \psi_{\alpha}(\bm{r}) & = & 
\frac{1-(-1)^{L+S+T}}{\sqrt{2}}\sum_{L'}
 {\cal Y}_{L'SJ}(\hat{\bm{r}})
 \ R_{L',L;S}^{J} (r)\ \eta_{T,T^3}\ ,\\
 {\cal Y}_{LSJ}(\hat{\bm{r}})&=&
\bigl[Y_L(\hat{\bm{r}})\otimes
 \chi_S \bigr]_{(J)} \ ,
\end{eqnarray}
where the two-nucleon
spin (isospin) wave function
is denoted by
$\chi_S$ ($\eta_{T}$).
The radial part of the above wave function is normalized,
in the plane wave limit, to be
\begin{eqnarray}
R_{L',L;S}^J(r) \rightarrow j_L(pr) \delta_{L,L'}.
\end{eqnarray}

With the multipole operators and the partial waves presented above,
I express the transition amplitude in terms of the partial
wave amplitudes:
\begin{eqnarray}
  T_{fi}  &=& 
\sum_{\alpha_i,m_i}\sum_{\alpha_f,m_f} \sum_{J_o,M_o}
  (-1)^{M_o} \  i^{J_o+L_i-L_f} 
\frac{(4\pi)^3}{\sqrt{2J_f+1}}
  Y^*_{L_i,m_i}(\hat{\bm{p}}_i)\;
  Y_{L_f,m_f}(\hat{\bm{p}}_f)  \nonumber\\
  && \times  
  (1/2,s_{1,i},1/2,s_{2,i}|S_i \mu_i)\;
(1/2,\tau_{1,i},1/2,\tau_{2,i}|T_i,T_i^3)\;
(L_i m_i S_i \mu_i |J_i M_i) 
\nonumber\\
  && \times  
  (1/2,s_{1,f},1/2,s_{2,f}|S_f \mu_f)\;
(1/2,\tau_{1,f},1/2,\tau_{2,f}|T_f,T_f^3)\;
(L_f m_f S_f \mu_f |J_f M_f) 
\nonumber\\
  && \times  
(J_i M_i J_o M_o  | J_f M_f)
\sum_{a=C,L} \braketa{T_a^{J_o}}\; Q_a^{J_o-M_o} \ ,
\end{eqnarray}
with the suffix $i$ ($f$) indicates the initial (final) state.
In the above equation, I used the abbreviation
\begin{eqnarray}
\eqn{simple-rme}
\braketa{T_a^{J_o}} & = & 
\bra{\psi_{\alpha_f}} |T_a^{J_o}|\ket{\psi_{\alpha_i}} \ ,
\end{eqnarray}
for the reduced matrix element defined by
\begin{eqnarray}
\bra{J_f, M_f}T_a^{J_o,M_o}\ket{J_i,M_i}
= \frac{1}{\sqrt{2J_f+1}} (J_i M_i J_o M_o | J_f M_f)\;
\bra{J_f}|T_a^{J_o}|\ket{J_i}\ .
\end{eqnarray}

\subsection{Cross section}\label{sec_cs2}

The unpolarized cross section for the $pp\to pn\pi^+$ reaction is 
\begin{eqnarray}
   \eqn{cs1}
d\sigma = \sum_{\bar{i},f} \frac{1}{v_{rel}}\frac{1}{(2\pi)^5}
\frac{1}{2\omega_\pi}
          \delta^{(4)}(P_i-q_\pi-P_f)
	   |T_{fi}|^2 
          d\bm{q}_\pi d\bm{p}_{1,f}d\bm{p}_{2,f}\ , 
\end{eqnarray}
where $\sum_{\bar{i},f}$ indicates the average (summation)
over the initial (final) spin and isospin states of the two nucleons.
The quantities $P_i$ and $P_f$ are the initial and final four total
momentum of the two-nucleon system; $P_\mu^2 = (P^0)^2 - \bm{P}^2$.
The relative velocity between the initial two nucleons is denoted by
$v_{rel}$.

I derive the pion angular distribution, retaining only partial
wave transition amplitudes of interest.
I am primarily interested in
the $^1S_0\to {}^3S_1$ transition amplitude,
$\bra{\psi_{{}^3\!S_1}}| T_a^{J_o}| \ket{\psi_{^1\!S_0}}$,
where the $\tilde{d}$ term plays an important role.
[It is noted that $\psi_{{}^3\!S_1}$ contains the $^3D_1$
component, as seen in \Eq{partial_wave}, and thus 
the $^1S_0\to {}^3D_1$ transition is also included in 
$\bra{\psi_{{}^3\!S_1}}| T_a^{J_o}| \ket{\psi_{^1\!S_0}}$.]
However, I also retain the $^1D_2\to {}^3S_1$ transition amplitude
for a later convenience.
I retain multipole operators with rank one ($J_o=1$) which dominantly
induce the transition.
I integrate over the final two nucleon momenta to obtain:
\begin{eqnarray}
   \eqn{cs2}
\frac{d\sigma}{d \Omega_\pi} = \int_0^{q_\pi^{\rm max}}\!\!\!\!\!\! d{q}_\pi \;
  \frac{E_p\sqrt{q^2_\pi+P_{\!\!\!f\; \mu}^{2}}\; p' q^2_\pi}{16\pi\; p\; \omega_\pi\; f_\pi^2}
|M|^2 \ ,
\end{eqnarray}
with
\begin{eqnarray}
|M|^2 &=& \frac{1}{4}\sum_{\alpha_i = {}^1\!S_0,{}^1\!D_2} | \bra{\psi_{{}^3\!S_1}}| T_C^{1} -
 q_\pi T_L^{1}| \ket{\psi_{\alpha_i}} |^2 
+ P_2(\cos\theta_\pi)
\left(
\frac{1}{4} | \bra{\psi_{^3\!S_1}}| T_C^{1} - q_\pi T_L^{1}| \ket{\psi_{^1\!D_2}}|^2\right.
\nonumber\\
&-&
\frac{1}{\sqrt{2}}
{\rm Re}\left[\bra{\psi_{^3\!S_1}}| T_C^{1} - q_\pi T_L^{1}| \ket{\psi_{^1\!D_2}} 
\bra{\psi_{^3\!S_1}}| T_C^{1} - q_\pi T_L^{1}| \ket{\psi_{^1\!S_0}}^*\right]
\biggr) \ .
\end{eqnarray}
The scattering angle of the pion with respect to the direction of the
initial proton is denoted by $\theta_\pi$, and $P_2 (x)$ is the
Legendre function of degree two.
The initial one nucleon energy is $E_p$ and
the maximum magnitude of the pion momentum is given by
\begin{eqnarray}
q_\pi^{\rm max}\!=\!\frac{
\sqrt{(E_p\!+m_N+m_\pi/2)(E_p\!+m_N-m_\pi/2)(E_p\!-m_N+m_\pi/2)(E_p\!-m_N-m_\pi/2)}}
{E_p} \ .
\end{eqnarray}
I will compare my calculation with 
``partial wave amplitudes'' extracted from experimental
data in Ref.~\cite{flammang}.
However, the ``partial wave amplitudes'' are actually different from 
the ordinary one, $\braketa{T_C^{J_o} - q_\pi T_L^{J_o}}$.
Therefore,
I establish the relation between them so that the comparison
makes sense, which will be done in the next paragraph.

In Ref.~\cite{flammang}, six lowest partial wave amplitudes are assumed to
contribute to the $pp\to pn\pi^+$ reaction in the energy region under
investigation, and are extracted from the data.
In extracting the amplitudes, they treated the final $NN$ system as the
``pseudo-deuteron'' which is the $NN$ scattering state with the relative
motion integrated over;
the pseudo-deuteron may have
the angular momentum different from the deuteron.
This means that 
they parameterized the data using a formulae
in which the $pp\to pn\pi^+$ reaction was regarded as
the $pp\to \mbox{``}d\mbox{''}\pi^+$ reaction; $\mbox{``}d\mbox{''}$
is the pseudo-deuteron.
More specifically, they parameterized the pion angular distribution
in the $pp\to pn\pi^+$ reaction using the formula:
\begin{eqnarray}
\eqn{cs_exp}
\frac{d\sigma}{d \Omega_\pi} = C_0 + C_2 P_2(\cos\theta_\pi) \ ,
\end{eqnarray}
with
\begin{eqnarray}
C_0 &=& \frac{|a_0|^2}{4} + \frac{|a_1|^2}{4} + \frac{|a_2|^2}{4} +
 \frac{|b_0|^2}{4} + \frac{|b_1|^2}{4} + \frac{|b_2|^2}{4} \ , \\
C_2 &=& \frac{|a_2|^2}{4} - \frac{{\rm Re} [a_0 a_2^*]}{\sqrt{2}} \ ,
\end{eqnarray}
where $a_0$ ($a_2$) is the partial wave amplitude for $^1S_0\to {}^3S_1$
($^1D_2\to {}^3S_1$); see Table IV of Ref.~\cite{flammang} for the other
partial wave amplitudes.
By comparing \Eq{cs_exp} with \Eq{cs2}, I can find the relation between
the two differential cross section formula.
To see the relation more clearly,
I denote the theoretical counterpart to $a_0$ ($a_2$) by $\tilde{a}_0$
($\tilde{a}_2$) and rewrite \Eq{cs2} as
\begin{eqnarray}
\eqn{cs3}
\frac{d\sigma}{d \Omega_\pi} = \frac{|\tilde{a}_0|^2}{4} +
\frac{|\tilde{a}_2|^2}{4} + \left(\frac{|\tilde{a}_2|^2}{4} 
- \frac{{\rm Re} [\tilde{\tilde{a}}_0 \tilde{\tilde{a}}^*_2]}{\sqrt{2}} \right)
P_2(\cos\theta_\pi) \ ,
\end{eqnarray}
with
\begin{eqnarray}
\eqn{tilde_a0}
|\tilde{a}_0|^2 &=& \int_0^{q_\pi^{\rm max}}\!\!\!\!\!\!\!\! d{q}_\pi 
  \frac{E_p\sqrt{q^2_\pi+P_{\!\!\!f\; \mu}^{2}}\; p' q^2_\pi}{16\pi\;
  p\; \omega_\pi\; f_\pi^2}
| \bra{\psi_{{}^3\!S_1}}| T_C^{1} - q_\pi T_L^{1}| \ket{\psi_{^1\!S_0}}
|^2 , \\
\eqn{tilde_a2}
|\tilde{a}_2|^2 &=& \int_0^{q_\pi^{\rm max}}\!\!\!\!\!\!\!\! d{q}_\pi 
  \frac{E_p\sqrt{q^2_\pi+P_{\!\!\!f\; \mu}^{2}}\; p' q^2_\pi}{16\pi\;
  p\; \omega_\pi\; f_\pi^2}
| \bra{\psi_{{}^3\!S_1}}| T_C^{1} - q_\pi T_L^{1}| \ket{\psi_{^1\!D_2}}
|^2 , \\
\eqn{tilde_a0a2}
{\rm Re} [\tilde{\tilde{a}}_0 \tilde{\tilde{a}}^*_2] &=&
\int_0^{q_\pi^{\rm max}}\!\!\!\!\!\!\!\! d{q}_\pi 
  \frac{E_p\sqrt{q^2_\pi+P_{\!\!\!f\; \mu}^{2}}\; p' q^2_\pi}{16\pi\;
  p\; \omega_\pi\; f_\pi^2}\nonumber\\
&\times&{\rm Re}\left[\bra{\psi_{^3\!S_1}}| T_C^{1} - q_\pi T_L^{1}| \ket{\psi_{^1\!D_2}} 
\bra{\psi_{^3\!S_1}}| T_C^{1} - q_\pi T_L^{1}|
\ket{\psi_{^1\!S_0}}^*\right] \ .
\end{eqnarray}
I distinctly used $\tilde{a}$ and $\tilde{\tilde{a}}$ because they are not
necessarily the same.\footnote{I use the symbol $a$ ($\tilde{a}$) in a
generic sense, referring to both ${a}_0$ and ${a}_2$ ($\tilde{a}_0$ and
$\tilde{a}_2$).
}
Which ($\tilde{a}$ or $\tilde{\tilde{a}}$) should be compared with
${a}_0$ and ${a}_2$ from the experimental data ?
I take the following way to find a solution.

At first, I factorize out the phase coming from the initial state
interaction as
\begin{eqnarray}
\tilde{a}_0 = e^{i\delta_0}\; \tilde{a}^\prime_0, \qquad
\tilde{a}_2 = e^{i\delta_2}\; \tilde{a}^\prime_2 \ ,
\end{eqnarray}
where $\delta_0$ ($\delta_2$) is the phase shift of the $^1\!S_0$
($^1\!D_2$) partial wave scattering. 
I choose $\tilde{a}^\prime_0$
and $\tilde{a}^\prime_2$ to be real.
The same factorization applies to $\tilde{\tilde{a}}_0$ and
$\tilde{\tilde{a}}_2$.
Then, as Step (i), I calculate $|\tilde{a}^\prime_0|$ and
$|\tilde{a}^\prime_2|$ from
\Eq{tilde_a0} and \Eq{tilde_a2}, respectively.
Next, as Step (ii), I solve a coupled equation
consisting of \Eq{tilde_a0a2} and 
$|\tilde{\tilde{a}}_0|^2 + |\tilde{\tilde{a}}_2|^2 =
|\tilde{a}_0|^2 + |\tilde{a}_2|^2$ [the r.h.s. is from Step (i)],
thereby finding a set of solutions, $\tilde{\tilde{a}}^\prime_0$
and $\tilde{\tilde{a}}^\prime_2$.
Step (ii) is closer to the way how $a_0$ and $a_2$ are extracted from the
data; each amplitude is extracted from the sum of the amplitudes (data).
My numerical result shows that the two sets of the solutions obtained
in Step (i) and (ii) are essentially the same in the absolute value.
[Step (i) cannot fix the sign.]\footnote{
Strictly speaking, $\tilde{a}$ obtained in my analysis is not
necessarily the same as $a$ from Ref.~\cite{flammang}.
In order to obtain $a$ theoretically, one calculates 
the analyzing power and the pion angular distribution
for the $pp\to pn\pi^+$ reaction,
taking account of all non-negligible partial wave amplitudes.
Then $a$ is extracted from those observables in the same way as
done in Ref.~\cite{flammang}.
However, the fact that $\tilde{a}$ obtained in Step (i) and (ii) are
essentially the same would indicate that my result would not change
drastically even if I took the ``ideal'' procedure to calculate $a$.
}
Therefore, I will employ the solution from Step (ii), in which the
relative phase between $\tilde{a}_0$ and $\tilde{a}_2$ can also be
fixed. 
(Hereafter, I do not distinguish between $\tilde{a}$
and $\tilde{\tilde{a}}$, and denote them by $\tilde{a}$.)
Still, the overall phase of $\tilde{a}$ has not been fixed.
In order to make the comparison with the data meaningful, one needs to
choose the phase convention for $\tilde{a}$ to be the same as that for
$a$.
I use $a_2$ and $\tilde{a}_2$ to match the phase conventions of the
experiment and the theory.
I calculate $\tilde{a}_2$
using the same operator used in
calculating $\tilde{a}_0$;
the $\tilde{d}$ term does not contribute here.~\footnote{
Because of the finite cutoff, the $\tilde{d}$ term gives a very small
contribution, which I safely neglect here.
}
I will fix the phase convention in this way in the next section,
followed by the comparison between $a_0$ and $\tilde{a}_0$.

\section{result}\label{sec_result}

\subsection{$^1D_2\to {}^3S_1$ transition amplitude}\label{sec_result_1d2}

\begin{figure}[t]
\begin{minipage}[t]{80mm}
\includegraphics[width=75mm]{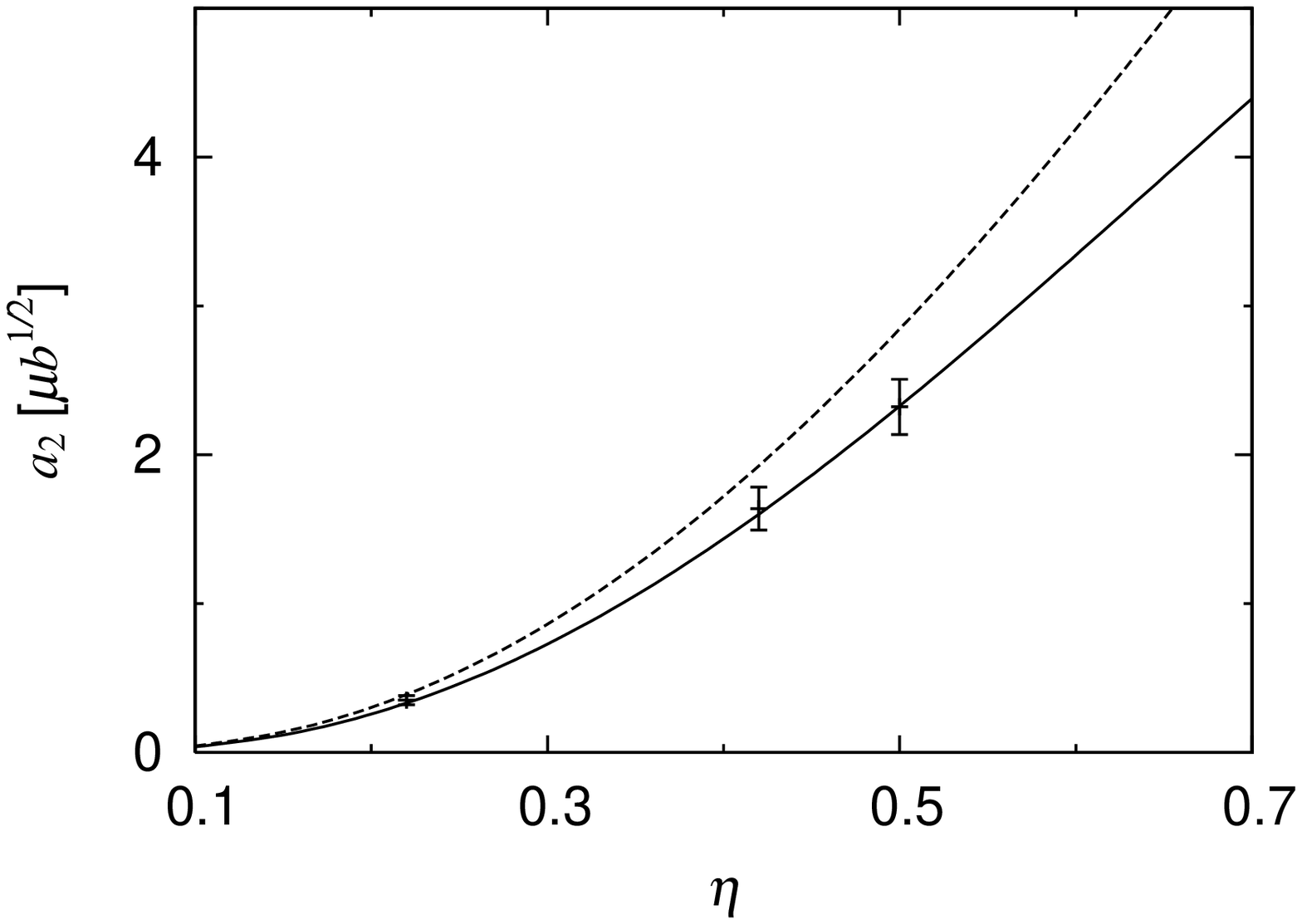}
\caption{\label{fig_a2_cdbonn}
The transition amplitude $\tilde{a}_2$ for $pp\to pn\pi^+$.
The chiral NLO $\pi$ production operator and the CD-Bonn $NN$ potential
are used.
The solid curve is obtained with $h_A =$ 2.10
while the dashed one with $h_A =$ 2.68.
Experimental data are from Ref.~\cite{flammang}.}
\end{minipage}
\hspace{2mm}
\begin{minipage}[t]{80mm}
\includegraphics[width=75mm]{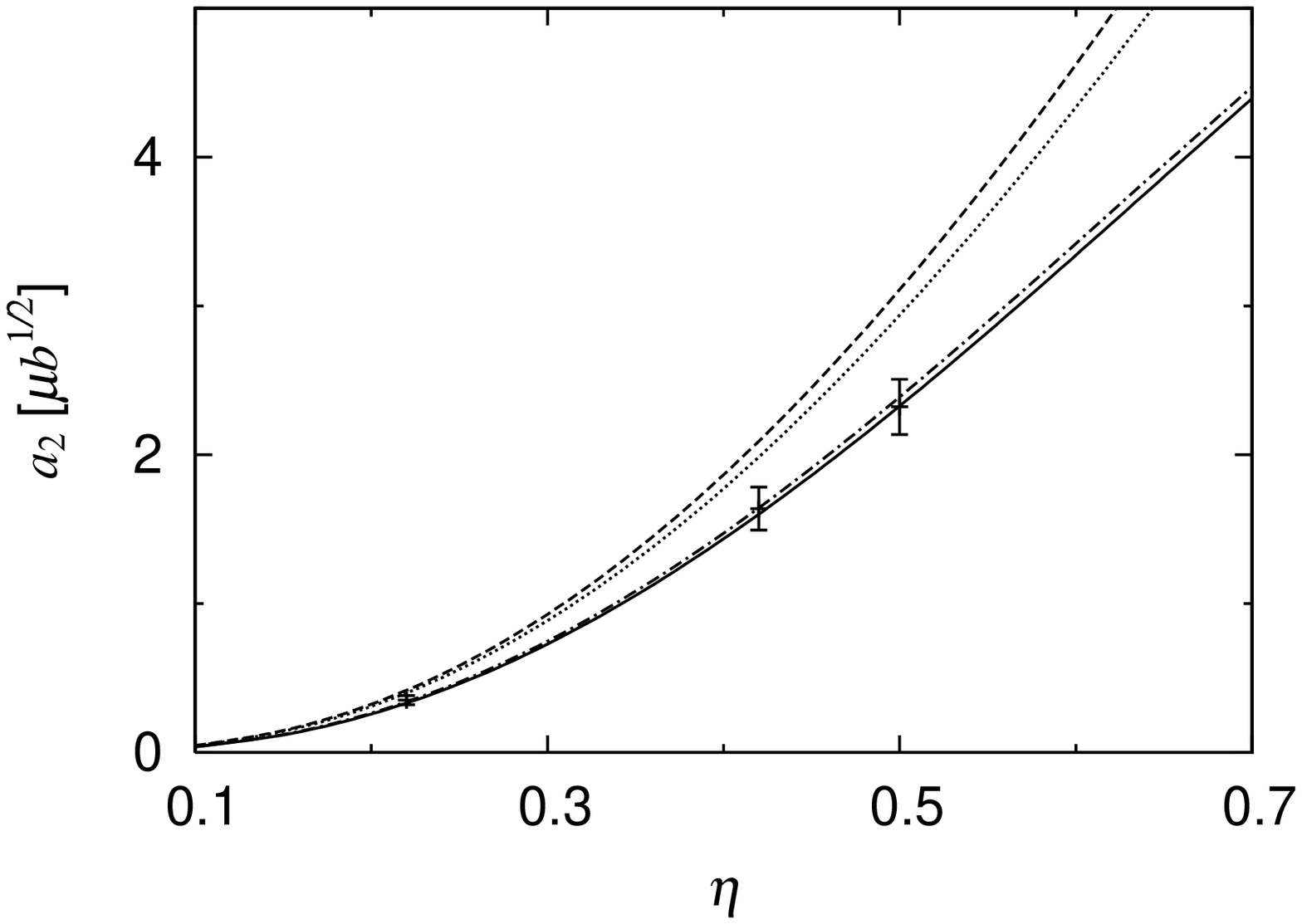}
\caption{\label{fig_a2_NN}
The transition amplitude $\tilde{a}_2$ for $pp\to pn\pi^+$ obtained with
the chiral NLO $\pi$ production operator.
I employ $h_A =$ 2.10.
The solid, dashed, dotted and dash-dotted curves are
 respectively $\tilde{a}_2$ obtained with
the CD-Bonn, AV18, Reid93, Nijmegen I $NN$ potentials.
The result with the N$^3$LO potential almost exactly falls on the
dash-dotted curve. Experimental data are from Ref.~\cite{flammang}.}
\end{minipage}
 \end{figure}

I calculate the 
$^1D_2\to {}^3S_1$ transition amplitude, $\tilde{a}_2$,
with the operators presented in Sec.~\ref{sec_op}.
Then, I compare $\tilde{a}_2$ with $a_2$ from the data\cite{flammang}.
In calculating $\tilde{a}_2$, I set the sharp cutoff $\Lambda = \infty$
because there is no counter term at this order which takes care of the
high momentum components of the operators integrated out.
In Fig.~\ref{fig_a2_cdbonn}, I show $\tilde{a}_2$ obtained with the
CD-Bonn $NN$ potential, as a function of
$\eta\; (\equiv q_\pi^{\rm max}/m_\pi)$.
As stated in Sec.~\ref{sec_di}, I consider neither
the Coulomb interaction nor isospin violation effects.
I use the CD-Bonn potential for the proton-neutron channel.
The solid curve is obtained with $h_A =$ 2.10 (and the corresponding
$c_i$'s) while the dashed curve with $h_A =$ 2.68.
I choose the overall sign for $\tilde{a}_2$ such that $a_2$ and
$\tilde{a}_2$ have the same sign, thereby fixing the phase convention.
The phase convention does not change if I use a finite cutoff such
as $\Lambda =$ 800~MeV in calculating $\tilde{a}_2$.
As observed in Fig.~\ref{fig_a2_cdbonn}, $\tilde{a}_2$ with $h_A =$ 2.10
falls exactly on the experimental data. 
We see the dependence of $\tilde{a}_2$ on the choice of the $NN$
potential in Fig.~\ref{fig_a2_NN}.
I used only $h_A =$ 2.10 in Fig.~\ref{fig_a2_NN}.
Although there is some dependence on the $NN$ potential, all
$\tilde{a}_2$ are on the vicinity of the data.
$\chi$PT gives a successful description for $a_2$ at this order.
Although $a_2$ is significantly larger than $a_0$ in magnitude, 
I stop the discussion on $a_2$ here and 
will give a detailed discussion 
on $a_0$ in the following subsections.
This is because I am primarily concerned with the validity of the
bridging program in this work, and $a_0$ is the one to be examined for
this purpose.

\subsection{$^1S_0\to {}^3S_1$ transition amplitude}\label{sec_result_1s0}

Now I move on to the $^1S_0\to {}^3\!S_1$ transition amplitude, 
$\tilde{a}_0$.
Because the relative phase between $\tilde{a}_2$ and $\tilde{a}_0$ is
fixed within the theory, and the overall phase has been fixed by
comparing $a_2$ and $\tilde{a}_2$, I am now able to compare $a_0$
and $\tilde{a}_0$ with the same phase convention.
It is noted that my phase convention for $\tilde{a}_0$ is the same as
the convention used in Ref.~\cite{hanhart_p-pi}.

At first, in order to see the importance of the $\Delta$,
I mention a result obtained with the $\pi$-production
operator without the $\Delta$.
I use the operators given in Eqs.~(\ref{eq:1b}), (\ref{eq:recoil-1b}),
(\ref{eq:c3})-(\ref{eq:d_i}) with the parameters taken from
Ref.~\cite{park}. 
The cutoff function is also the same as that used in Ref.~\cite{park}
(the Gaussian cutoff). 
I found a very large cutoff dependence of $\tilde{a}_0$
calculated with the $\Delta$-less $\chi$PT. 
Depending on the cutoff ($\Lambda_\chi =$ 500, 600 and 800~MeV), 
the contributions from the two-body operators to
$\tilde{a}_0$ are different by a factor of 4,
which demonstrates the failure of the $\Delta$-less $\chi$PT in
describing the $p$-wave $\pi$-production.
I note that
this result is quite different from the situation of the low-energy weak
process where the $\Delta$-less $\chi$PT gives the cross sections with a
small cutoff dependence.

\begin{figure}[t]
\begin{minipage}[t]{80mm}
\includegraphics[width=75mm]{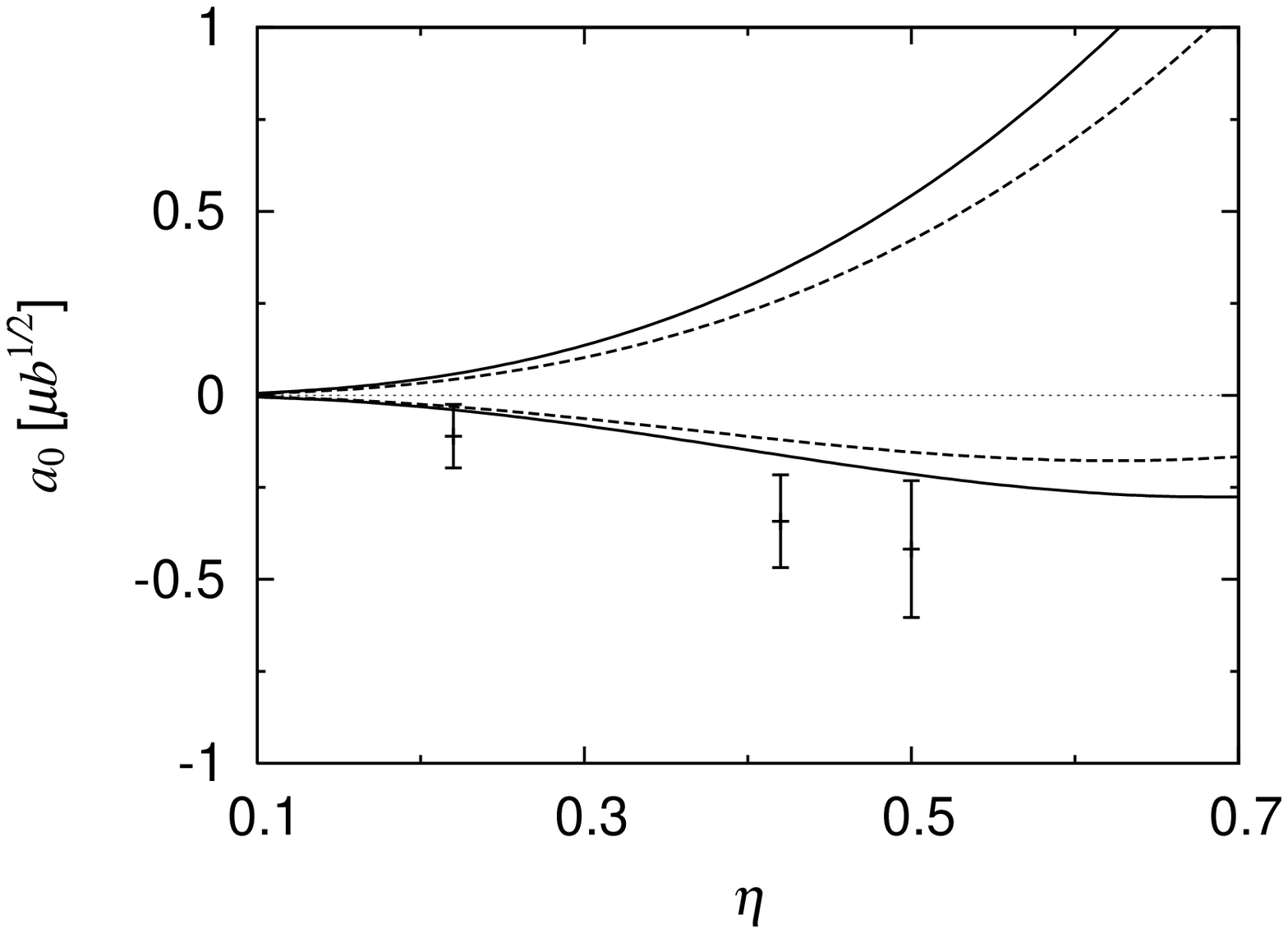}
\caption{\label{fig_cdbonn_ha}
The transition amplitude $\tilde{a}_0$ for $pp\to pn\pi^+$.
The chiral NLO $\pi$ production operator and the CD-Bonn $NN$ potential
are used.
The solid and dashed curves correspond to
$h_A =$ 2.10 and 2.68, respectively; $\Lambda =$ 800~MeV.
The upper (lower) curves are obtained with (without) the
$\tilde{d}$ term.
Experimental data are from Ref.~\cite{flammang}.}
\end{minipage}
\hspace{2mm}
\begin{minipage}[t]{80mm}
\includegraphics[width=75mm]{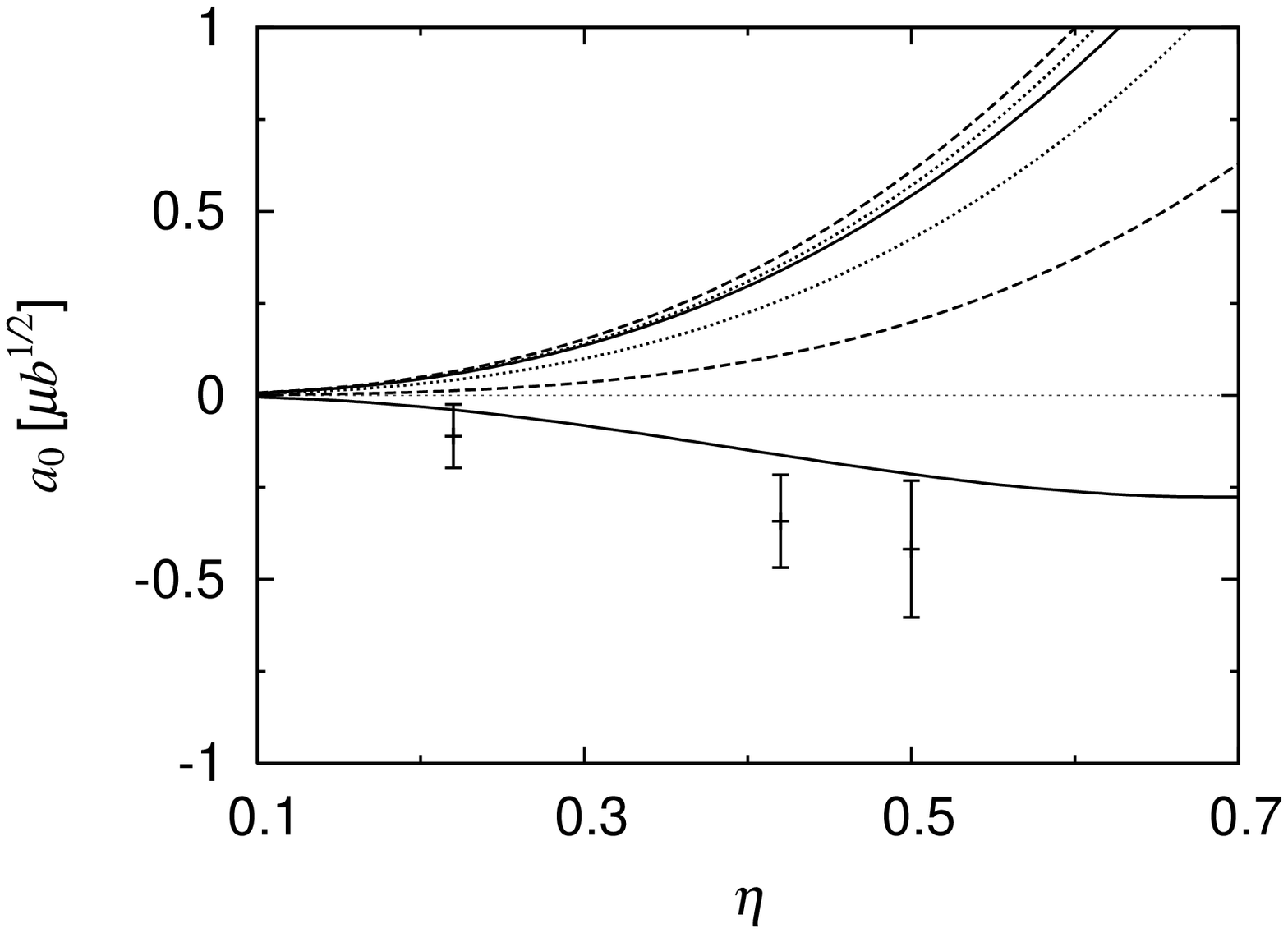}
\caption{\label{fig_cdbonn_lambda}
The transition amplitude $\tilde{a}_0$ for $pp\to pn\pi^+$.
The CD-Bonn $NN$ potential is used.
The solid, dashed and dotted curves correspond to
$\Lambda =$ 800, 600 and 500~MeV, 
respectively; $h_A$ = 2.10.
The upper (lower) three curves are obtained with (without) the
$\tilde{d}$ term.
Experimental data are from Ref.~\cite{flammang}.}
\end{minipage}
\end{figure}

Next I present results obtained with the operator including the
$\Delta$.
The operators have been presented in Sec.~\ref{sec_op}, and the LEC
$\tilde{d}$ has been fixed using the low-energy weak process
in Sec.~\ref{sec_di}.
I am interested in how reasonably and reliably one can predict $\tilde{a}_0$
for the $pp\to pn\pi^+$ reaction using this $\chi$PT-based operator.
For this purpose, I examine the dependence of $\tilde{a}_0$ on several
inputs, $\Lambda$, $h_A$, $C_2^{N\Delta}$, and the $NN$ potential.
At first, in Fig.~\ref{fig_cdbonn_ha},
I present a result obtained with the CD-Bonn $NN$ potential and two
choices of the $h_A$ value.
The partial wave amplitude $\tilde{a}_0$ is rather different from the
experimental counterpart, $a_0$.
For comparison, I also show a result obtained without the $\tilde{d}$
term.
This result with $\tilde{d} = 0$ is similar to the case with 
$\delta = 0$ in Fig.~3 of Ref.~\cite{hanhart_p-pi} where
a negative $\tilde{d}$ value 
({\it e.g.}, $\delta = -0.2$) brings the theoretical amplitude into
the agreement with the experimental one.
In my calculation, however, the positive $\tilde{d}$ value (see the last row
of the second and third columns in Table~\ref{tab_d}) separates
$\tilde{a}_0$ and $a_0$
furthermore; even the sign of $\tilde{d}$ fixed by the low-energy weak
process is inconsistent with the experimental data of the $\pi$-production.
I change the values of $\Lambda$ and $h_A$ and examine the dependence
of $\tilde{a}_0$ on these inputs.
The cutoff dependence is shown in Fig.~\ref{fig_cdbonn_lambda}.
The situation of the disagreement does not change.~\footnote{
For some combinations of $h_A$, $\Lambda$ and $NN$ potential,
the sign of $\tilde{d}$ is consistent with the data.
However, the strength is not enough to bring $\tilde{a}_0$ into
agreement with $a_0$.
}
This result shows that the bridging program among reactions with quite
different kinematics is not necessarily successful.
This is understandable if we recall another case.
For example, a chiral nuclear force\cite{n3lo,epel} describes the
elastic $NN$ scattering over a fairly wide energy region, partly because
the LECs in it have been fixed using data from the same energy region.
Therefore, it is no wonder to find that the operator fixed in the
low-energy process cannot describe well the intermediate-energy process.
In order to accurately describe the two reactions in the different
energy regions simultaneously, data from both of the energy regions
would be necessary to fix the LECs.
It is also expected that higher order terms are necessary to accurately
reproduce the data from the wide energy region, as in the
case of the nuclear force.

We find from Fig.~\ref{fig_cdbonn_lambda} the cutoff dependence
($\sim$10\% level for $\tilde{a}_0$; the $\tilde{d}$ term included) which is
much reduced compared with the $\Delta$-less case.
We also find a certain amount of
dependence on the choice of $h_A$ (Fig.~\ref{fig_cdbonn_ha}), even if
the $\tilde{d}$ value has been
adjusted to eliminate the dependence on it at the low-energy kinematics.
\footnote{\label{foot1}The variation of $h_A$ should be compensated for by the
change of $c_i$ and other higher order one-pion rescattering diagrams;
not by the change of $\tilde{d}$.
I will come back to this point later.}
This means that the $\Delta$ operator and the $\tilde{d}$ contact
operator have quite different dependence on the kinematics,
and it is
important to take care of each component of the operator individually.
The dependence on the kinematics is also found in
Table~\ref{tab_kin} where I tabulated contributions from each component
of the operator to $\tilde{a}_0$ for two kinematics; one for
low-energy weak process ($pp\to de^-\nu_e$) where $\tilde{d}$ is fixed, 
and the other for the $\pi$-production ($pp\to d\pi^+$, $\eta =$ 0.5).
\begin{table}[t]
\renewcommand{\arraystretch}{1.2}
\tabcolsep=2.3mm
\caption{\label{tab_kin}
Contribution from each component of the operator to the matrix element for
$pp\to d\pi^+$ (second row) and $pp\to de^-\nu_e$ (third row).
Each contribution is divided by the ``sum'', and therefore, ``1B''
 actually gives the matrix elements with the same sign for both cases.
The CD-Bonn potential, $h_A =$ 2.10, and $\Lambda =$ 800~MeV are used.
Some operators, not shown here, contribute to the $\pi$
production by a small amount; at most, ``WT'' contributes by $\sim 0.08$.
}
 \begin{tabular}[t]{cccccccc}\hline
kinematics       &1B     &$\Delta\pi$&$\Delta$CT&$c_3$   &$c_4$   &$\tilde{d}$&sum (arb.units)\\\hline
$pp\to d\pi^+$   &$-$0.60& $-$0.54 &     0.97&$-$1.11 & 0.71   & 1.57    &$-$0.0251\\
$pp\to de^-\nu_e$&0.996  &   0.009 & $-$0.008&  0.015 &   0.005&$-$0.017 &    0.757\\\hline
 \end{tabular}
\end{table}
This situation is in contrast to the bridging program done in
Ref.~\cite{park}. 
In Ref.~\cite{park}, the operator fixed by a low-energy weak process
({\it i.e.}, the tritium $\beta$-decay) was used in another weak process
which takes place in a relatively similar kinematics, and the result was
given with a small cutoff dependence.
I might say that it is not important to take care of each
component of the operator individually in this case. 
Rather, one needs to take care of the sum of each component.
This is also a reason why the pionless theory, 
whose two-body operator is just a contact operator,
can reasonably describe several weak processes in a low-energy
region\cite{piless}.

From Table~\ref{tab_kin}, we also find that the contributions from the
NLO operators is comparable to those from the LO operator.
The LO contributions mostly cancel each other.
For this reaction,
there is no sign of the convergence in the chiral expansion
up to this order.
Furthermore, there is also a significant cancellation among
the NLO contributions, which may make the amplitude sensitive to the
higher order contributions.
This situation reminds us of the result in Ref.~\cite{ando3,kim} where
the $pp\to pp\pi^0$ reaction near threshold was studied with $\chi$PT.
They found that some higher order two-pion-exchange mechanism contributes
more than lower order one-pion-exchange mechanism, leading to a poor
convergence of the chiral expansion.
It would be important to do a higher order calculation of
the $p$-wave $\pi$ production to see the
convergence of the chiral expansion.

We find in Table~\ref{tab_kin} that the contribution from the
$\tilde{d}$ term is substantial.
This result may also indicate the importance of a higher order calculation, 
which I will argue in the following.
The large contribution from the $\tilde{d}$ term means that
its dependence on the kinematics is rather influential on
$\tilde{a}_0$. 
Because I am working at the NLO, two-pion-exchange (TPE) mechanisms
are not explicitly considered but mimicked by the $\tilde{d}$ term.
If the TPE and the $\tilde{d}$ term have rather different dependence on
the kinematics, $\tilde{a}_0$ given by the NLO and NNLO calculations may be
significantly different.

\begin{figure}[t]
\begin{minipage}[t]{80mm}
 \includegraphics[width=75mm]{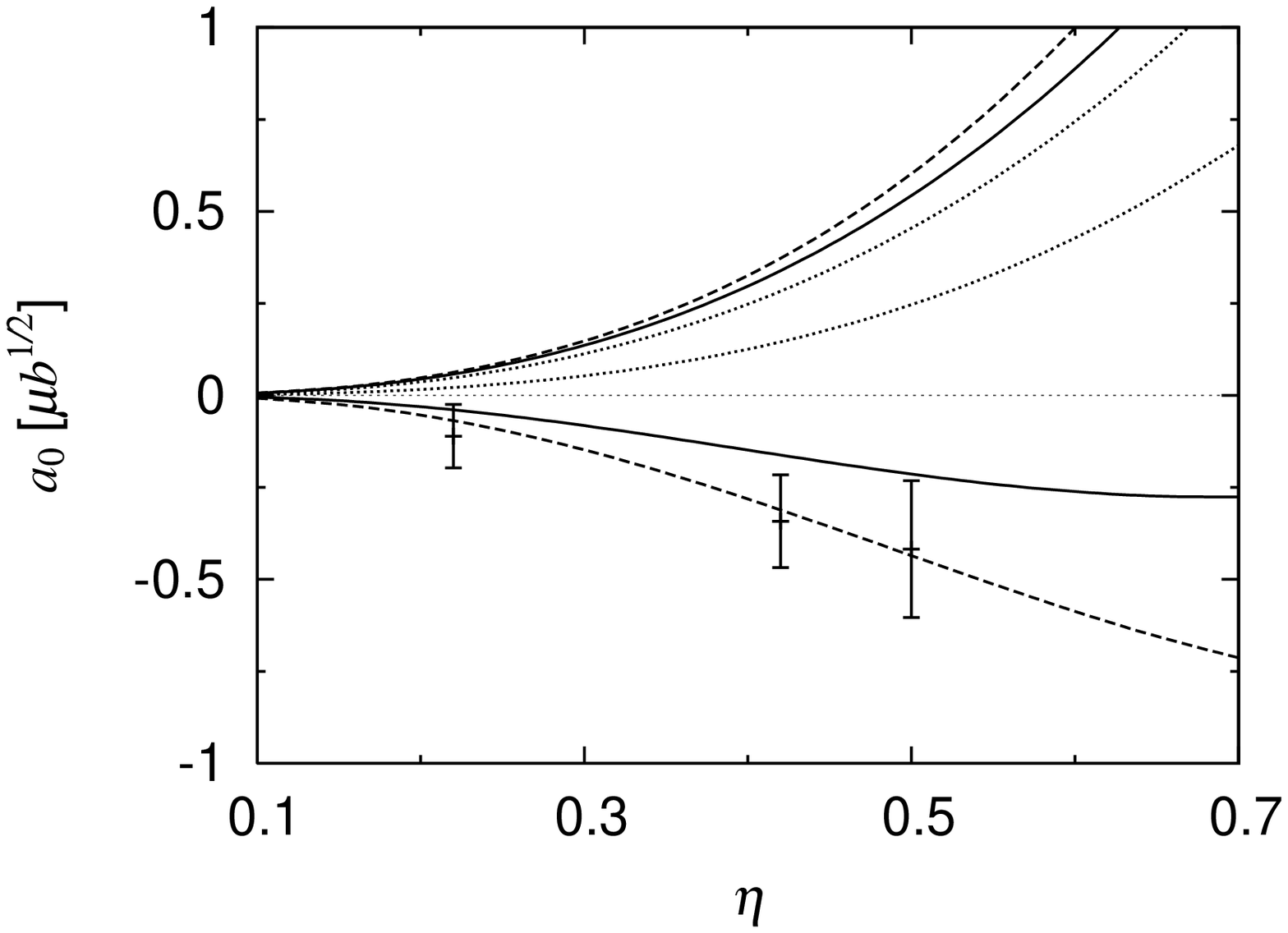}
 \caption{\label{fig_a0_nn}
 The transition amplitude $\tilde{a}_0$ for $pp\to pn\pi^+$.
The solid, dashed and dotted curves are respectively obtained with
the CD-Bonn, AV18 and N$^3$LO $NN$ potentials; $h_A$ = 2.10 and
$\Lambda$ = 800~MeV.
The upper (lower) three curves are obtained with (without) the
$\tilde{d}$ term.
Data are from Ref.~\cite{flammang}.}
\end{minipage}
\hspace{2mm}
\begin{minipage}[t]{80mm}
 \includegraphics[width=75mm]{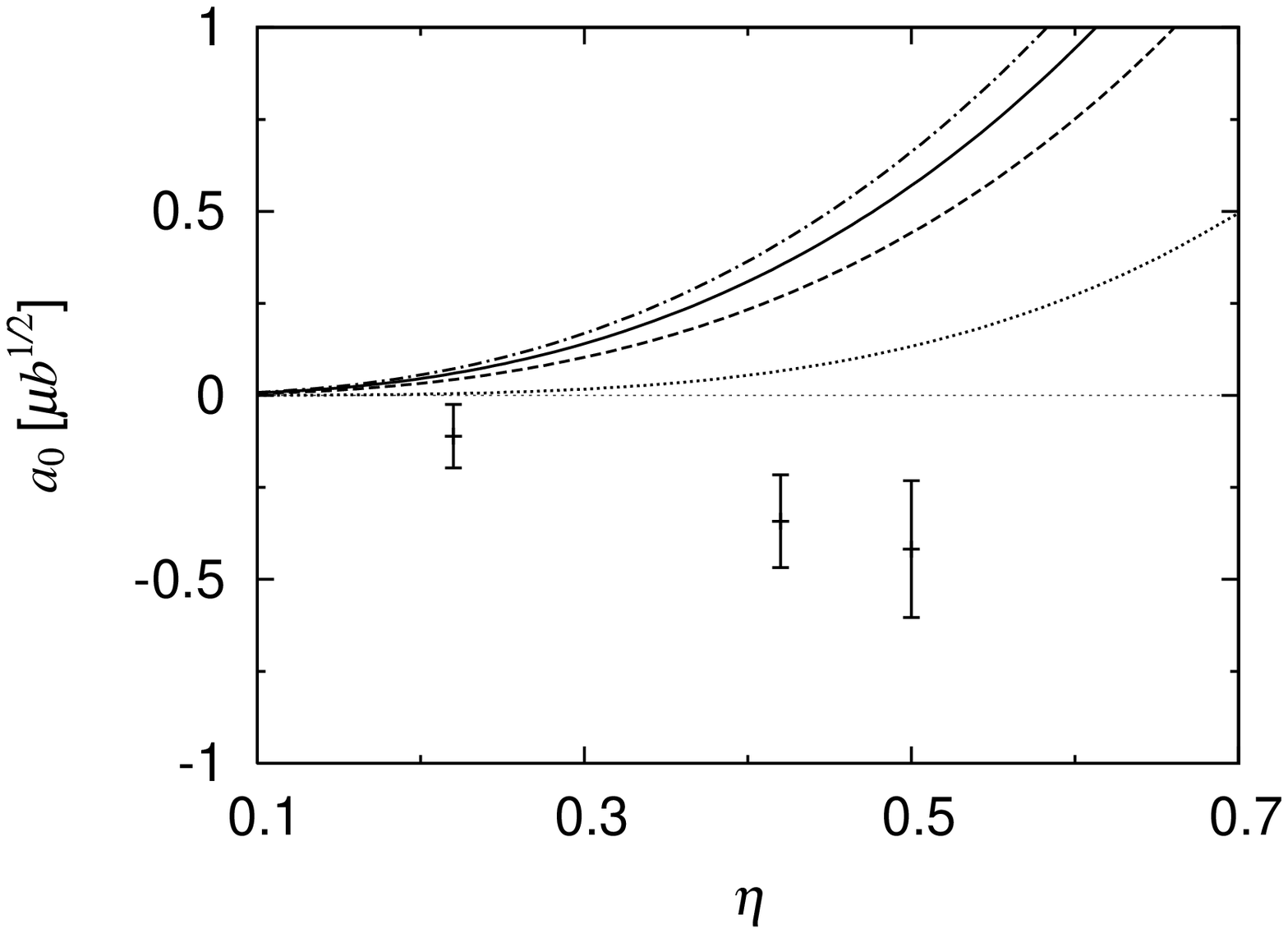}
 \caption{\label{fig_a0_nn_500-2}
 The transition amplitude $\tilde{a}_0$ for $pp\to pn\pi^+$.
The solid, dashed, dotted and dash-dotted curves are respectively
 obtained with
the CD-Bonn, AV18, Reid93 and Nijmegen I $NN$ potentials; $h_A$ = 2.10
 and $\Lambda$ = 500~MeV.
The N$^3$LO potential almost gives the solid curve.
The $\tilde{d}$ term is included.
Data are from Ref.~\cite{flammang}.}
\end{minipage}
\end{figure}

I show results obtained with various $NN$ potentials to see the
dependence of $\tilde{a}_0$ on it.
The result is shown in 
Fig.~\ref{fig_a0_nn}.
We find a considerable dependence on the $NN$ potential, even though all
$NN$ potentials give almost the same amplitude in the low-energy region
where the $\tilde{d}$ value has been fixed.
The $\tilde{d}$ term is quite sensitive to the short-distance behavior
of the wave function because of its point-like nature.
Therefore, when the cutoff is relatively large and thus
the short-distance behavior of the wave function
is very dependent on the $NN$ potential,
the $\tilde{d}$ value is also largely dependent on the $NN$ potential.
In case of $\Lambda = 800$~MeV, for example, 
a part of the dependence of $\tilde{a}_0$ on the $NN$ potential
 is ascribable to that
the $\tilde{d}$ term alone is too simple to compensate for
the difference in the short distance physics.
However, I consider that there is a more influential source of the
dependence on the $NN$ potential, because we still find the dependence
in the case of $\Lambda = 500$~MeV (Fig.~\ref{fig_a0_nn_500-2}). 
Let me explain more on this in the following.
In the rest of this paragraph, I discuss the case with $\Lambda = 500$~MeV. 
At this cutoff, the details of the short distance physics have been
integrated out substantially, and therefore there is no significant
difference among the wave functions for different $NN$ potentials any
more.
Recall that low-momentum $NN$ potentials obtained from various
phenomenological $NN$ potentials are very similar at
$\Lambda\sim 400$~MeV\cite{vlowk,rg1}.
In this situation, the $\tilde{d}$ values for different $NN$ potentials
should be almost the same to give almost the same $\tilde{a}_0$.
Contrary to this expectation, I obtained quite different $\tilde{d}$ 
as seen in the second row of TABLE~\ref{tab_d}, leading to rather
different $\tilde{a}_0$.
\footnote{\label{unique}In the next subsection where I include one more
contact term, we will see that the expectation is realized.}
I suspect that
the $\tilde{d}$ term alone is too simple to simulate
the operator {\it to be simulated},~\footnote{
Even though we do not know, there should exist 
an operator which the $\tilde{d}$ term tries to simulate.
The operator should include all non-negligible chiral operators other than
those explicitly considered already.
I refer to the operator as the operator {\it to be simulated}.
}
and that
my procedure of determining $\tilde{d}$,
discussed in the previous section, 
gives the $\tilde{d}$ term
which is far from being as approximate to 
the operator {\it to be simulated} as it can be.
I suspect that this is the main source of the $NN$ dependence of
$\tilde{a}_0$.
The relatively large $h_A$-dependence of $\tilde{d}$ (and $\tilde{a}_0$)
is also likely to have the same origin.
In order to improve the situation,
one would need to do a higher order calculation, and include a
few more contact operators so that a better simulation can be
done.~\footnote{
One might suspect that the dependence on the $NN$ potential
is due to a difference in the phase shift of the $^1S_0$ partial wave
in the energy region above the pion production threshold.
Among the $NN$ potentials I use, 
the phase shifts from the Reid93 and the N$^3$LO potentials are
noticeably different from the others at this energy.
However,
I do not consider the difference in the phase shift to be influential
because
the inclusion of one more contact term results in that
all of the $NN$ potentials with $\Lambda = 500$~MeV
give essentially the same $\tilde{a}_0$ over
the entire energy region under consideration.
Note that the contact term does not reflect the difference in the long
range behavior (phase shift) of the wave function.
}

Because I treated the contact $NN\to N\Delta$ interaction in the
phenomenological manner, as discussed in Sec.~\ref{sec_op}, it is
informative to study an impact of changing
the coupling on the amplitude $\tilde{a}_0$.
For this purpose, I change the $C_2^{N\Delta}$ value by $\pm$ 10\%,
re-fit the $\tilde{d}$ value in the way discussed in Sec.~\ref{sec_di},
and calculate $\tilde{a}_0$.
I found that $\tilde{a}_0$ is less dependent on the
variation of $C_2^{N\Delta}$ than that of $h_A$.
This result may be understood as follows.
The contact-induced $\Delta$-excitation mechanism
[Fig.~\ref{fig_diag}(c)] is similar to the $\tilde{d}$ term
[Fig.~\ref{fig_diag}(e)] in the sense that
Fig.~\ref{fig_diag}(c) is reduced to Fig.~\ref{fig_diag}(e)
in the $\Delta$-less theory.
Therefore, one may expect that
the variation of $C_2^{N\Delta}$ is fairly well compensated for by the
change of $\tilde{d}$.

\subsection{$^1S_0\to {}^3S_1$ transition amplitude with one more
  counter term}\label{sec_result_1s0_2}

As seen in the previous section, the bridging program was not
successful; the $\chi$PT-based operator with 
$\hat{d} (\equiv m_N f_\pi^2 \tilde{d})$ fixed by the
low-energy weak process does not reproduce
the partial wave amplitude for the $\pi$-production, $a_0$,
extracted from the data.
Because I have pointed out several reasons for expecting a higher
order calculation, here I try to see what happens there by doing a
simple extension of my calculation.
The extension is to add a higher order counter term to the NLO
$\chi$PT-based operator used in the previous section.
I use the following counter term of 
${\cal O}[(m_\pi/m_N)^2]$~\footnote{There exist
${\cal O}[(m_\pi/m_N)^{3/2}]$ operators coming from pion-loop diagrams.
For a simple analysis, I use
the ${\cal O}[(m_\pi/m_N)^2]$ counter term.
}:
\begin{eqnarray}
\eqn{l_ct2}
&& {\cal L}_{\rm CT}^{(2)} 
        =        -\frac{\hat{e}}{m_N f_{\pi}^2\Lambda^2} 
        N^{\dagger}\bm{\tau}\cdot\vec{\sigma}\cdot\vec{\nabla}\bm{\pi} N\,
        \left(N^{\dagger} \vec{\nabla}^2N + {\rm h.c.}\right) \ ,
\end{eqnarray}
where the dimensionless LEC is denoted by $\hat{e}$.
This is not a general form of the counter terms at this order,
and there are other
counter terms with different spin-isospin and derivative structures.
However, it is sufficient to consider only this counter term for my
purpose of gaining a rough insight into a higher order calculation.

Now I have the two independent LECs: $\hat{d}$ and $\hat{e}$.
I fix these two LECs so that the following two conditions are
satisfied.
The first condition is the same as that used in  Sec.~\ref{sec_di} for
fixing $\hat{d}$. This condition is from the low-energy weak process.
The second condition is due to $a_0$ extracted from 
the experimental data for the $pp\to pn\pi^+$ reaction\cite{flammang}.
I choose the LECs so that the central value of the empirical amplitude,
$a_0 (\eta = 0.5) = -0.418\; \mu{\rm b}^{1/2}$, is reproduced.
As seen in the previous section, one cannot reproduce 
the $\pi$-production partial wave amplitude ($a_0$) 
by using $\hat{d}$ fitted to the amplitude of the low-energy weak
process alone.
The disagreement between $a_0$ and $\tilde{a}_0$ is rather serious;
even the sign of $\hat{d}$ is inconsistent with the data in some cases.
Therefore, it is not obviously expected that the addition of the $\hat{e}$
term with a {\it natural} strength brings $\tilde{a}_0$ into agreement
with $a_0$.
However, such a set of $\hat{d}$ and $\hat{e}$ does exist, as presented in
Table~\ref{tab_de}.
\begin{table}[t]
\renewcommand{\arraystretch}{1.2}
\tabcolsep=2.3mm
\caption{\label{tab_de}
Dimensionless contact couplings, $\hat{d}$ and $\hat{e}$.
The first column is the sharp cutoff value, and the others are values of
 the LECs.
The 2--4th rows are for the $\hat{d}$ values while 5--7th rows for the
 $\hat{e}$ values.
For each $NN$ potential, the
 left side is the LECs for $h_A =$ 2.10 while the right side for
$h_A =$ 2.68.
}
 \begin{tabular}[t]{cccccccccccc}\hline
 $\Lambda$ &\multicolumn{2}{c}{CD-Bonn}&\multicolumn{2}{c}{AV18}&\multicolumn{2}{c}{Reid93}
 &\multicolumn{2}{c}{Nij I}&\multicolumn{2}{c}{N$^3$LO}\\\hline
500& 0.57& 0.57& 0.45& 0.46& 0.43& 0.44& 0.51& 0.51&$-$0.22&$-$0.21\\
600& 0.49& 0.48& 0.20& 0.19& 0.15& 0.14& 0.37& 0.36&$-$4.12&$-$4.08\\
800&$-$1.15&$-$1.34& 3.06& 3.30& 3.12& 3.35&$-$5.41&$-$6.14& 8.42& 8.35\\\hline
500& 0.85& 0.85& 0.94& 0.94& 0.96& 0.95& 0.90& 0.90& 1.44& 1.43\\
600& 1.10& 1.11& 1.35& 1.36& 1.40& 1.41& 1.20& 1.22& 5.15& 5.12\\
800& 3.39& 3.60&$-$1.55&$-$1.84&$-$1.62&$-$1.89& 8.38& 9.23&$-$7.83&$-$7.75\\\hline
 \end{tabular}
\end{table}

\begin{figure}[t]
\begin{minipage}[t]{80mm}
\includegraphics[width=75mm]{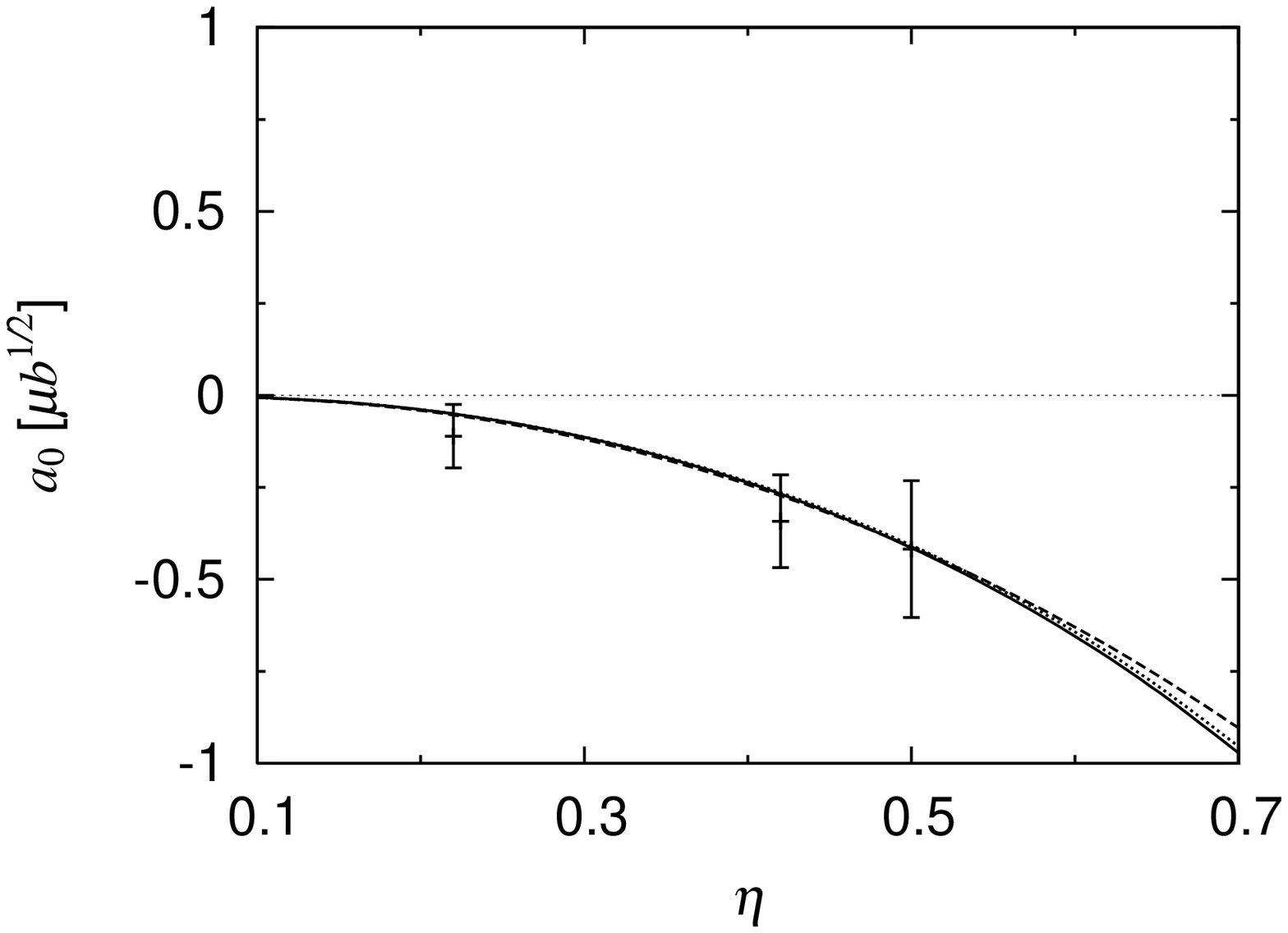}
\caption{\label{fig_a0_d2_cdbonn}
The transition amplitude $\tilde{a}_0$ for $pp\to pn\pi^+$.
The chiral NLO $\pi$ production operator 
plus the counter term [\Eq{l_ct2}]
and the CD-Bonn $NN$ potential are used.
The solid, dashed and dotted curves correspond to
$\Lambda =$ 800, 600 and 500~MeV, respectively, and $h_A =$ 2.10.
Experimental data are from Ref.~\cite{flammang}.}
\end{minipage}
\hspace{2mm}
\begin{minipage}[t]{80mm}
\includegraphics[width=75mm]{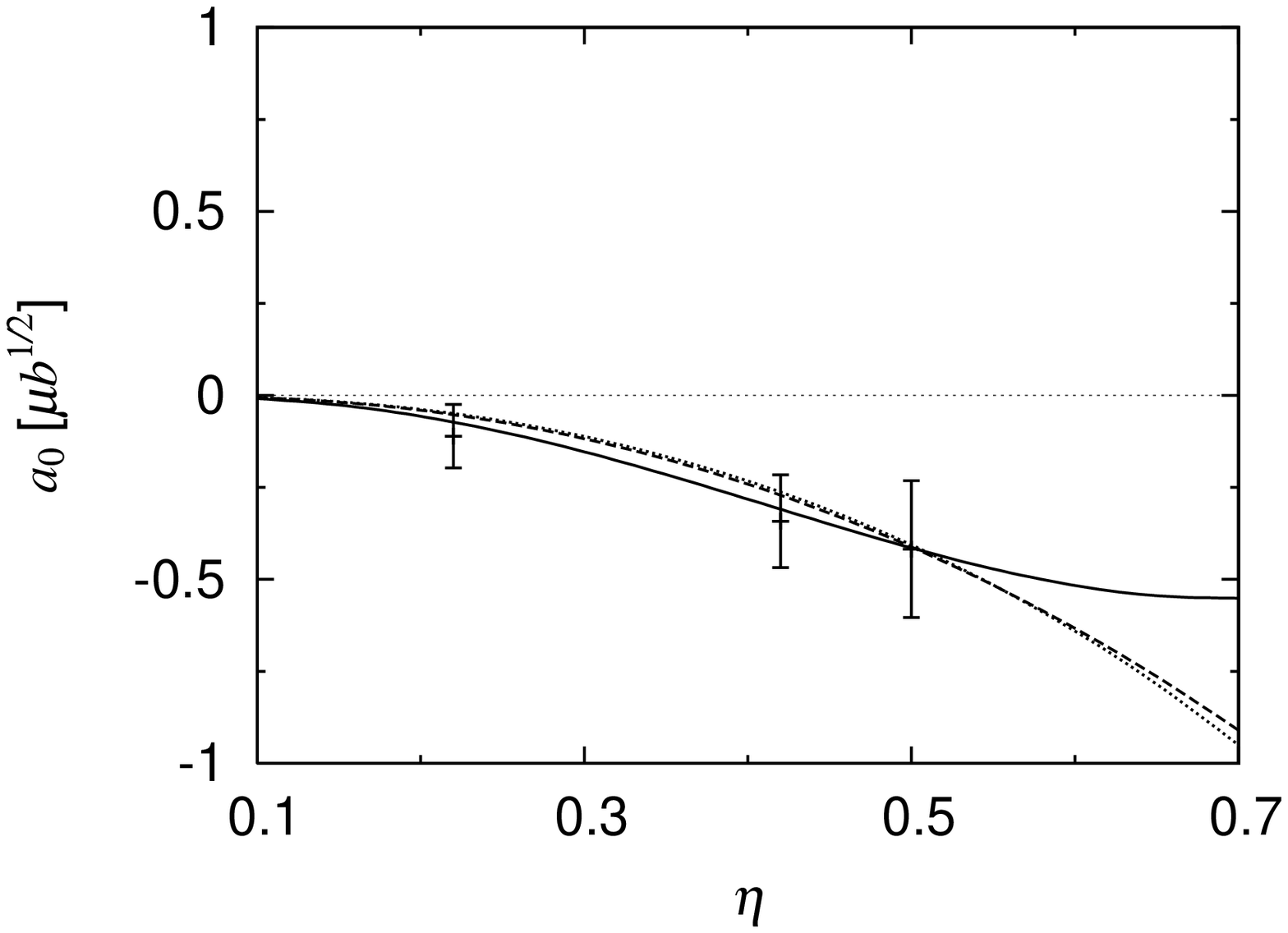}
\caption{\label{fig_a0_d2_av18}
The transition amplitude $\tilde{a}_0$ for $pp\to pn\pi^+$.
The AV18 $NN$ potential is used.
The other features are the same as Fig.~\ref{fig_a0_d2_cdbonn}.
}
\end{minipage}
\end{figure}

I calculate $\tilde{a}_0$ 
with the parameter sets in Table~\ref{tab_de}.
The results for the CD-Bonn and the AV18 $NN$ potentials
are shown in
Figs.~\ref{fig_a0_d2_cdbonn} and \ref{fig_a0_d2_av18}, respectively.
I show results obtained with $h_A =$ 2.10 only. The dependence of 
$\tilde{a}_0$ on $h_A$ is small in this case.
By construction, the calculated $\tilde{a}_0$ goes through the central
value of the data at $\eta =$ 0.5.
The $\eta$ dependence of $\tilde{a}_0$ is consistent with the data for
all $NN$ potentials and for all $\Lambda$.
Particularly, all $NN$ potentials 
give essentially the same $\tilde{a}_0$ for $\Lambda = 500$~MeV.
From Table~\ref{tab_de}, we also observe that the couplings $\hat{d}$ and
$\hat{e}$ for $\Lambda = 500$~MeV are similar for all the $NN$ potentials,
except for the N$^3$LO potential.
In fact, this result has been expected; see footnote \ref{unique}.
For $\Lambda = 800$~MeV, a relatively large dependence of $\tilde{a}_0$
(two types of the behavior) on the $NN$
potential is observed, within the consistency with the data.
Regarding the CD-Bonn and the Nijmegen I potentials, 
$\tilde{a}_0$ ($\Lambda = 800$~MeV) is very
similar to $\tilde{a}_0$ ($\Lambda = 500$ or 600~MeV);
see Fig.~\ref{fig_a0_d2_cdbonn}.
On the other hand, 
the other $NN$ potentials give 
$\tilde{a}_0$ ($\Lambda = 800$~MeV) whose
$\eta$-dependence is quite different from
$\tilde{a}_0$ ($\Lambda = 500$ or 600~MeV);
see Fig.~\ref{fig_a0_d2_av18}.
From a viewpoint of the renormalization group, the operators with
$\Lambda = 500$, 600 and 800~MeV should be related by integrating out the high
momentum states whose effects are simulated by the
renormalization of $\hat{d}$ and $\hat{e}$.
If this relation 
is realized, these operators should
give the same observables, to the extent that the contact operators
simulate the high energy modes integrated out.
From this viewpoint, the results for the CD-Bonn
(Fig.~\ref{fig_a0_d2_cdbonn}) 
and the Nijmegen I 
are understandable,
while the others (Fig.~\ref{fig_a0_d2_av18})
are not.
However, the situation may change in a correct higher order calculation
where the TPE is explicitly considered.
This is because
the model space employed here probably
has a resolution at which the contact operators cannot
accurately simulate 
the intermediate-range mechanism such as TPE.
If the TPE is explicitly considered, then 
the contact operators do not
have to mimic the intermediate-range mechanism, and more accurately
describe the shorter-range mechanism.

\begin{figure}[t]
\begin{minipage}[t]{80mm}
\includegraphics[width=75mm]{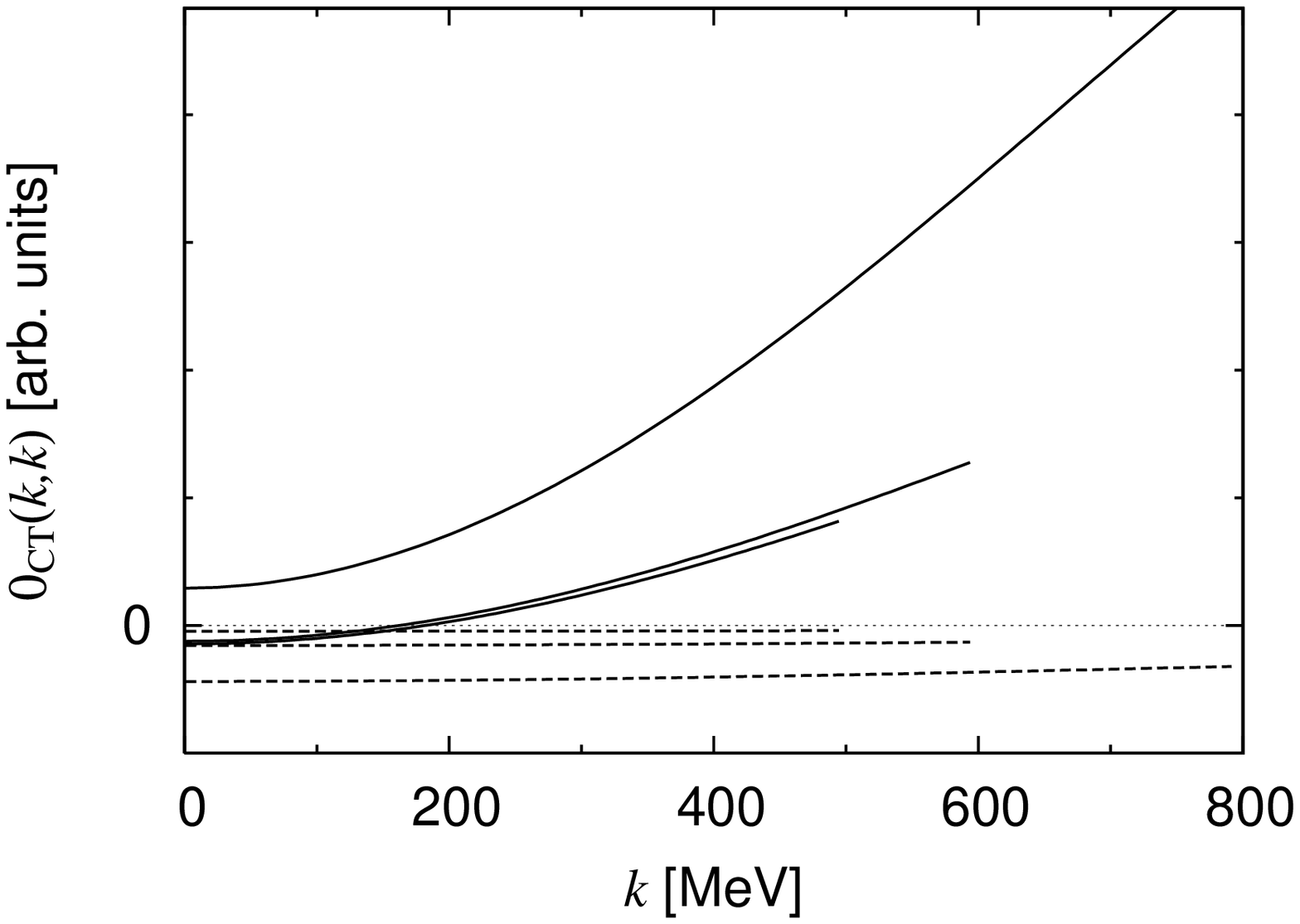}
\caption{\label{fig_ct_cdbonn}
The contact operators for the CD-Bonn potential. 
The diagonal components in the momentum space are given.
The solid curves are
 parameterized by both $\hat{d}$ and $\hat{e}$, while the dashed curves
 by $\hat{d}$ only. The $k$-coordinate value at the end point of a
 curve indicates the value of $\Lambda$ for the operator. $h_A$ = 2.10.
}
\end{minipage}
\hspace{2mm}
\begin{minipage}[t]{80mm}
\includegraphics[width=75mm]{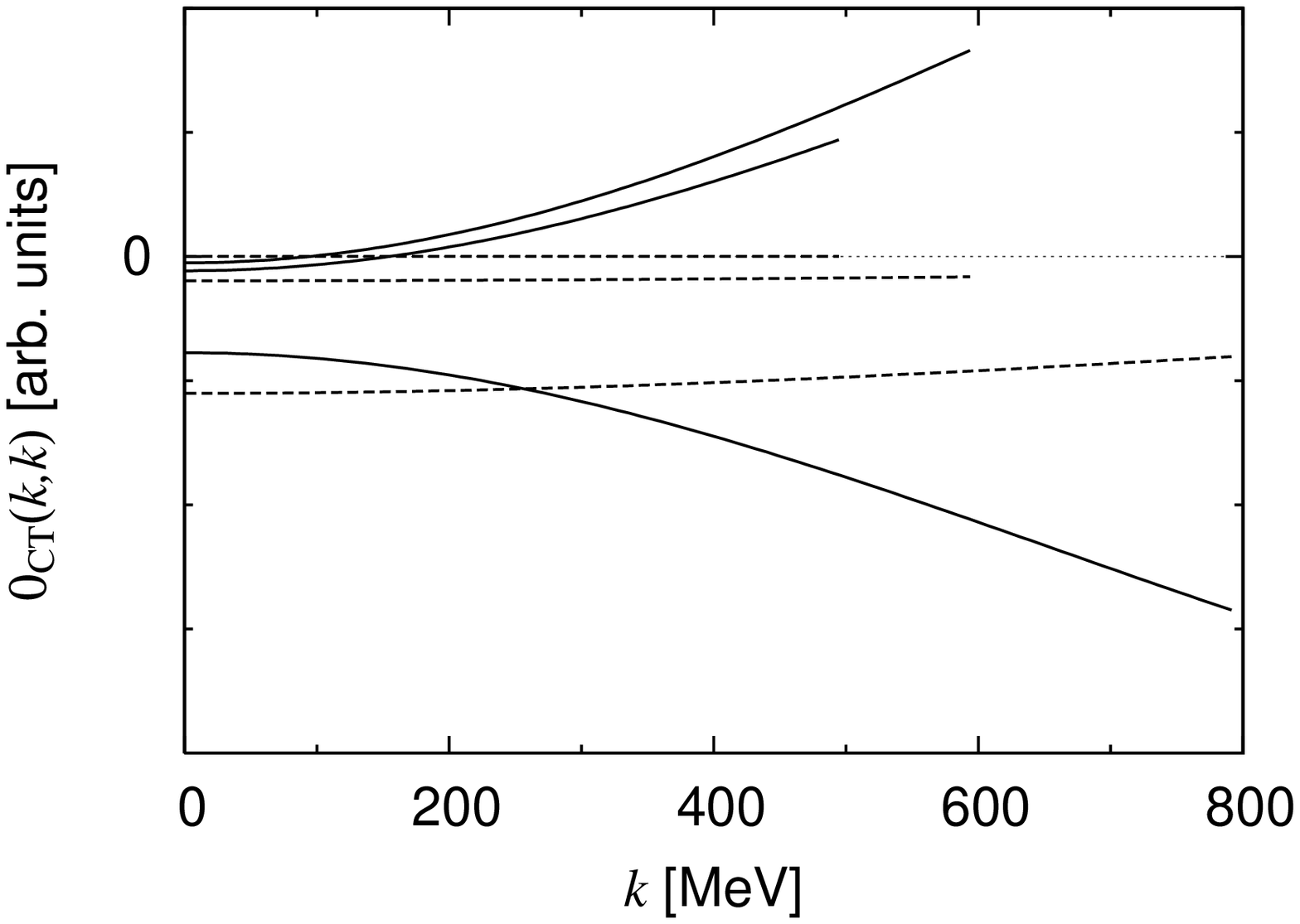}
\caption{\label{fig_ct_av18}
The contact operators for the AV18 potential. The other features are the
 same as Fig.~\ref{fig_ct_cdbonn}. The scale is also the same as
Fig.~\ref{fig_ct_cdbonn}.
}
\end{minipage}
\end{figure}

\begin{figure}[t]
\begin{minipage}[t]{80mm}
\includegraphics[width=75mm]{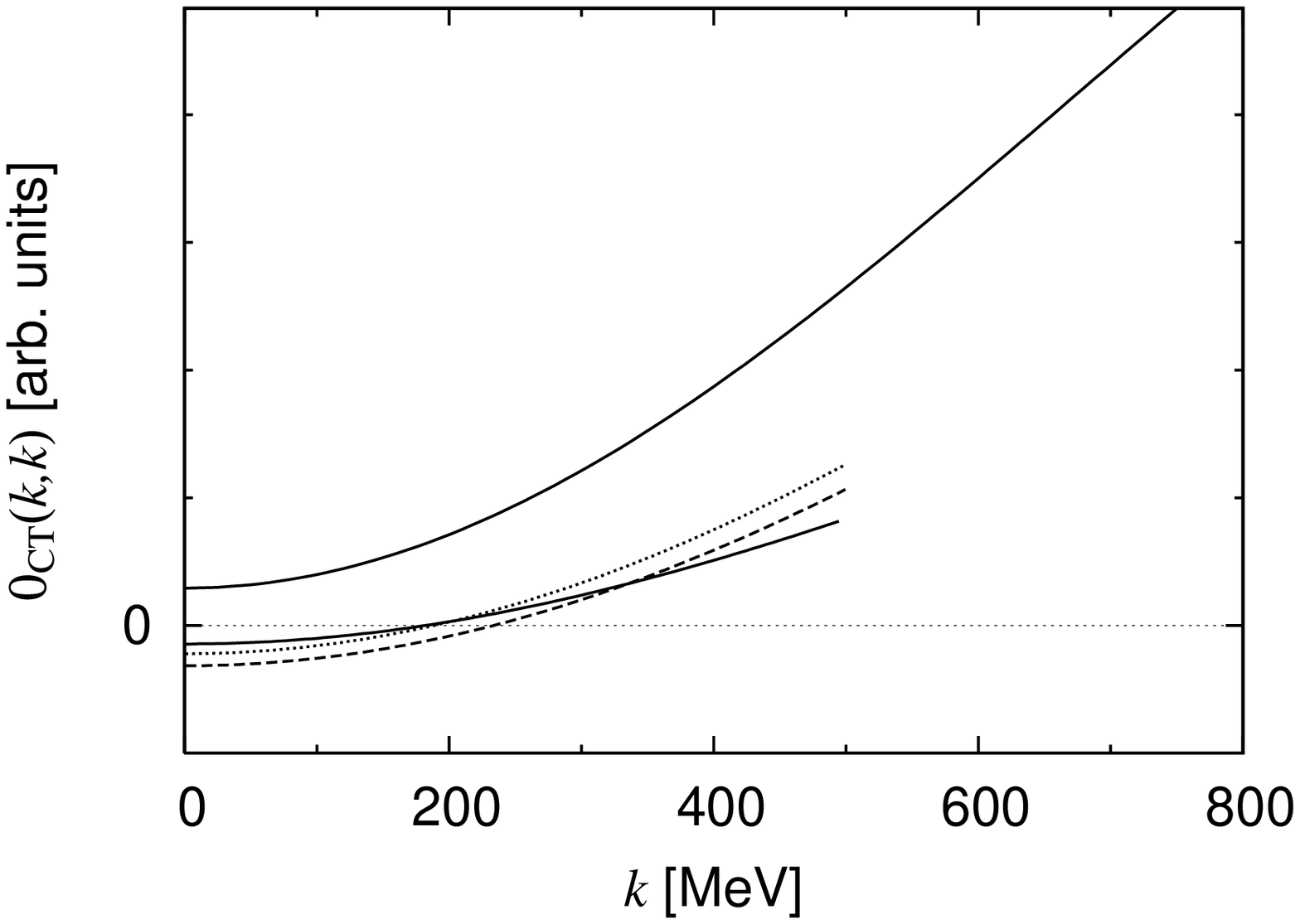}
\caption{\label{fig_rg_cdbonn}
RG running of contact operators for the CD-Bonn potential. 
The solid curves are the same as those in Fig.~\ref{fig_ct_cdbonn}.
(The curve for $\Lambda = 600$~MeV is not shown.)
The dashed and dotted curves are derived from the solid curve for
$\Lambda = 800$~MeV by solving the RG equation.
The dashed (dotted) curve is obtained with the kinematics for the
$pp\to d\pi^+$ ($pp\to de^+\nu_e$) reaction.
}
\end{minipage}
\hspace{2mm}
\begin{minipage}[t]{80mm}
\includegraphics[width=75mm]{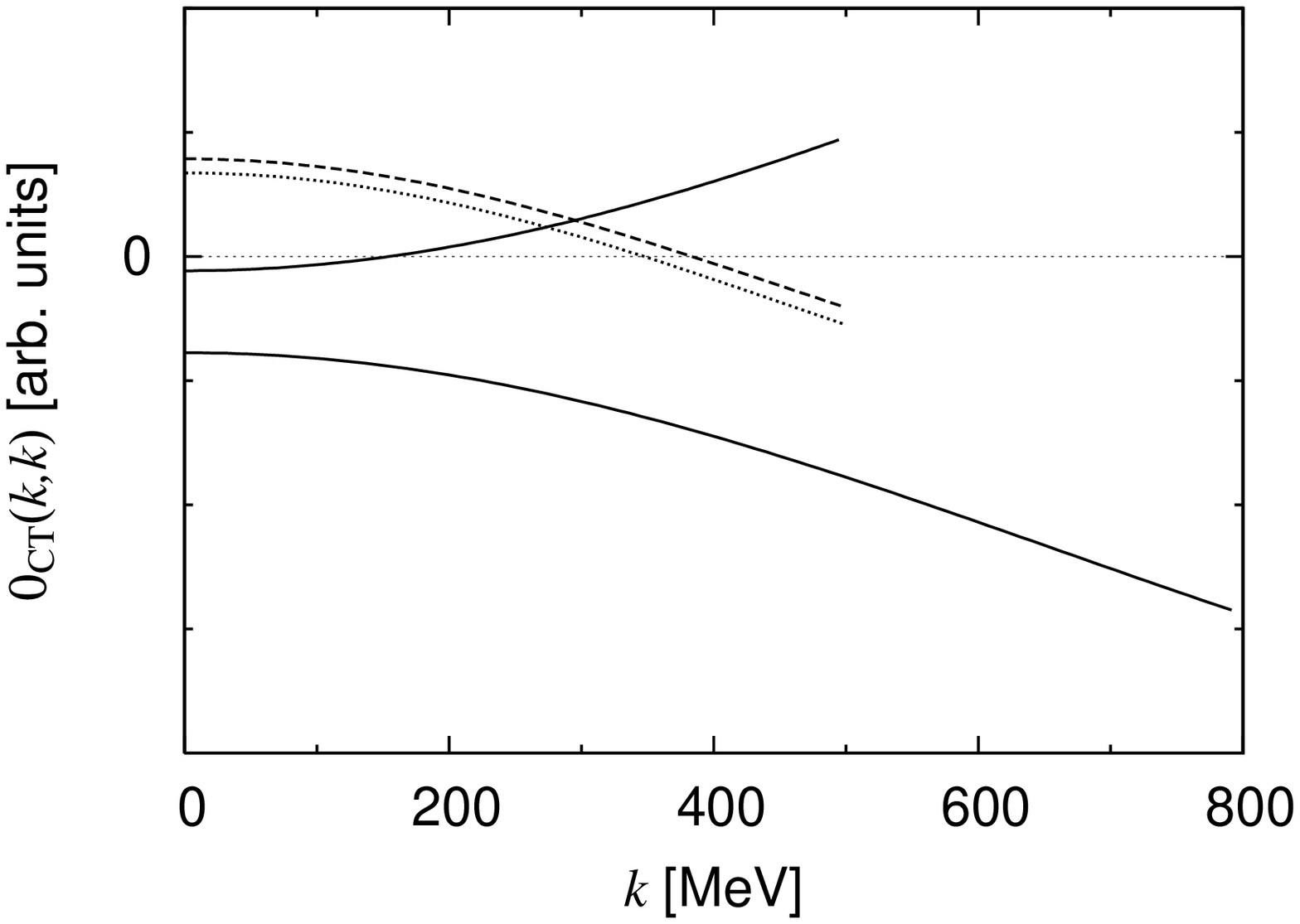}
\caption{\label{fig_rg_av18}
RG running of contact operators for the AV18 potential. 
The solid curves are the same as those in Fig.~\ref{fig_ct_av18}.
The other features are the same as Fig.~\ref{fig_rg_cdbonn}.
}
\end{minipage}
\end{figure}

I look into the contact operators for different $NN$ potentials,
and understand the similar (different)
$\eta$-dependence of $\tilde{a}_0$ for
$\Lambda$ = 500 (800)~MeV among different $NN$ potentials.
As representatives, I show the contact operators for the CD-Bonn and the
AV18 potentials in Figs.~\ref{fig_ct_cdbonn} and \ref{fig_ct_av18},
respectively.
We can find that the contact operators, 
including both $\hat{d}$ and $\hat{e}$,
for $\Lambda$ = 500 (800)~MeV are very
similar (different), leading to the
quite similar (different) $\eta$ dependence of $\tilde{a}_0$.
Starting with the operators for $\Lambda = 800$~MeV, one can reduce
$\Lambda$ to examine the running of the operators using the Wilsonian RG
equation.
I refer the readers to Refs.~\cite{NA,NA2} for a detailed explanation of
how I reduce $\Lambda$, and just show the result here.
In Fig.~\ref{fig_rg_cdbonn}, the RG running of the contact operator for
the CD-Bonn potential is given. 
For a comparison, the contact operator
for $\Lambda = 500$~MeV, whose LECs are given in Table~\ref{tab_de},
are also shown.
We can see that the contact operators ($\Lambda = 500$~MeV)
derived in the different two ways, one
from directly fitting to the data and the other from the RG running, are
fairly similar.
Although the figure shows the result for the diagonal components of the
operators, the same level of the agreement is confirmed for off-diagonal
components.
For the AV18 potential, however, the two contact operators 
($\Lambda=500$~MeV) with the
different origins are quite different as shown in Fig.~\ref{fig_rg_av18}.
This result has been expected by observing Fig.~\ref{fig_a0_d2_av18}
where the contact operators with $\Lambda = 500$ and 800 MeV do not give
the same result, indicating that the two operators are not equivalent.

I compare the contact operator including both $\hat{d}$ and
$\hat{e}$ (Table~\ref{tab_de}), and those with only $\hat{d}$
(Table~\ref{tab_d}) in Figs.~\ref{fig_ct_cdbonn} and \ref{fig_ct_av18}.
A naive expectation is that a contact operator with $\hat{d}$
only (a dashed curve) is an approximation of the one with $\hat{d}$
and $\hat{e}$ (the solid curve).
In the two figures, this expectation is not always the case,
and therefore we have to be careful about the
convergence of the chiral expansion of the short distance physics.
In any case, the $\hat{d}$ term alone cannot be a good approximation of
the $\hat{d}$ plus $\hat{e}$ terms and, for that matter, 
not a good approximation of the operator {\it to be simulated}.
It is recalled that this observation has been used 
in the previous subsection to argue over a reason for
the dependence of $\tilde{a}_0$ on $h_A$ and the $NN$ potential.
By including the $\hat{e}$ term, the situation is much improved
in this point.
We found the very small dependence of $\tilde{a}_0$ 
on the choice of $h_A$ and
on the $NN$ potential
for $\Lambda = 500$~MeV.
Regarding the $\hat{d}$ and $\hat{e}$ values,
as seen in Table~\ref{tab_de}, 
they hardly depend on $h_A$.
This is quite consistent with the discussion given in 
footnote~\ref{foot1}.

\section{summary}\label{sec_summary}

I determined the LEC $\hat{d}$ using the low-energy weak process and
then used it to predict 
the partial wave amplitude, $\tilde{a}_0$ ($^1S_0\to {}^3S_1$), for
the $pp\to pn\pi^+$ reaction.
Through this work, I tried to explore the power of $\chi$PT
that enables one to bridge different reactions.
My investigation is more stringent test of this aspect of $\chi$PT than
similar analyses in the literature
because the reactions connected through $\chi$PT here take place under
significantly different kinematics.

I started with the chiral Lagrangian including the nucleon, pion and
$\Delta$.
It is mandatory to include the $\Delta$ explicitly for
describing the $p$-wave $\pi$ production.
With the Lagrangian, I constructed, up to NLO of the chiral expansion 
proposed in Ref.~\cite{hanhart_p-pi,mod_counting},
a set of operators which describes
the Gamow-Teller transition in low-energy weak processes and the
$p$-wave $\pi$-production.
I fixed the remaining unknown LEC $\hat{d}$ (indirectly) using the
experimental data of the low-energy weak process.
Then I calculated the partial wave amplitudes ($\tilde{a}$) for the
$pp\to pn\pi^+$ reaction. 
I chose the phase convention such that $\tilde{a}_2$ 
($^1D_2\to {}^3S_1$) has the same phase as $a_2$ extracted from the data
in Ref.~\cite{flammang}.
My prediction of $\tilde{a}_0$ using the NLO operator does not agree
with the data.
I used several different sets of the inputs such as the $NN$ potential,
$h_A$ and $\Lambda$.
Even though there is some dependence of $\tilde{a}_0$ on these inputs,
all results differ from the data in the similar manner.
Unfortunately, even the sign of $\hat{d}$, fixed by the weak process,
is sometimes not consistent with the data.
This result indicates that the bridging program between 
the two reactions with quite different kinematics is not always successful.
In the literature, we have sometimes found an argument which supposes
that the bridging program works.
Given the result here, it is clear that we need to seriously study a
feasibility of the bridging program, particularly for reactions with
different kinematics.
This conclusion may be disappointing, but still understandable if we recall
the success of the chiral nuclear force.
The chiral nuclear force accurately describes the $NN$
scattering over a wide energy region, partly because the LECs included
have been fixed using data from the same energy region.
In fact, there are several reasons to expect a
higher order calculation to significantly improve the situation.
First of all, one may naively think that the NLO, tree level, operator
is too simple to describe the $\pi$ production.
We know that two-pion-exchange mechanism
and higher order contact terms are necessary
for accurately describing the $NN$ elastic
scattering near the $\pi$ production threshold.
Second, the $\hat{d}$ term largely contributes to $\tilde{a}_0$, and
therefore it would be important to extract the TPE mechanism from the
$\hat{d}$ term, thereby describing the intermediate-range mechanism more
elaborately and reducing the role of the $\hat{d}$ term.
Third, $\tilde{a}_0$ is considerably dependent on the choice of the
$NN$ potential, which  means that the single contact term is too simple to
cancel out different short distance physics for different $NN$ potentials.
I also argued that a substantial part of the $NN$-dependence 
(and $h_A$-dependence) of
$\tilde{a}_0$ is likely to be ascribable to the fact that the $\hat{d}$ term
alone is too simple to simulate the operator {\it to be simulated}.
Meanwhile, a higher order calculation is also desirable to see the convergence
of the chiral expansion; regarding the $pp\to pn\pi^+$ reaction,
we found no sign of the convergence up to this order.

In order to explore, even roughly, a result of a higher order
calculation, I added a higher order counter term, with the LEC
$\hat{e}$, to the NLO operator.
I fitted the LECs $\hat{d}$ and $\hat{e}$ to both the Gamow-Teller
amplitude for the low-energy weak
process and $\tilde{a}_0$ for the $\pi$ production.
I found a set of the LECs with the
natural strength.
The LECs are mostly independent of the choice of $h_A$ as they should;
without the $\hat{e}$ term, however, $\hat{d}$ is rather dependent on $h_A$.
With this parameter set, $\eta$-dependence of $\tilde{a}_0$ is described
in a way consistent with experimental data, irrespective of the choices
of the $NN$
potential, $h_A$ and $\Lambda$.
This result would be an indication that a higher order calculation is
promising.
I found that the single $\hat{d}$ term is not always a good approximation
of the $\hat{d}$ plus $\hat{e}$ terms.
I also showed using the RG analysis that
the contact operators with different cutoff are not 
always equivalent.
These findings tell us to be careful about the convergence of the chiral expansion,
and also the importance of going to a higher order calculation.

\begin{acknowledgments}
I acknowledge B. K. Jennings for reading the manuscript
 and comments.
I also thank
C. Hanhart and D. R. Phillips for useful discussions.
The Natural Sciences and Engineering Research Council of Canada is
 thanked for financial support. TRIUMF receives federal funding via a
 contribution agreement through the National Research Council of Canada.
\end{acknowledgments}

\begin{appendix}

\section{multipole expansion of $O_{\Delta\pi}$ operator}
I present our calculational procedure for the $O_{\Delta\pi}$ operator
[\Eq{pi-d}].
I start with the Fourier transform of $O_{\Delta\pi}$:
\begin{eqnarray}
\eqn{fourier_pi-d}
&&\int \frac{dk^3}{(2\pi)^3} e^{-i\bm{k}\cdot\bm{r}}
 \frac{\bm{k}\; \bm{\sigma}_2\cdot\bm{k}}{m_\pi^{\prime 2} + k^2}\ 
\frac{f(k)
}{m_\Delta - m_N - p_o^2/m_N +
  (\bm{p}^\prime+\bm{q}_\pi/2)^2/2\mu} \ ,
\end{eqnarray}
where I only consider one term in $O_{\Delta\pi}$; the other terms and
a constant factor are omitted.
The function $f(k)$ is a cutoff function:
$f(k) = \exp\left(-k^2/\Lambda_G^2\right)$ with $\Lambda_G = $ 2~GeV.
In this equation, I expand the energy denominator as follows:
\begin{eqnarray}
\eqn{exp_deno}
D(\bm{p}^\prime,\bm{q}_\pi) &=& \frac{1}{m_\Delta - m_N - p_o^2/m_N +
  (\bm{p}^\prime+\bm{q}_\pi/2)^2/2\mu}  \nonumber\\
&=& 4\pi \sum_{\ell} (-1)^\ell \sqrt{2\ell + 1} \left[Y_{\ell}(\hat{\bm{p}}^\prime)\otimes 
Y_{\ell}(\hat{\bm{q}}_\pi)\right]_{(0)}^0 D_\ell(p^\prime,q_\pi) \ ,
\end{eqnarray}
where $D_\ell(p^\prime,q_\pi)$ is defined by
\begin{eqnarray}
D_\ell(p^\prime,q_\pi)
&=& \frac{1}{2}\int^1_{-1} D(\bm{p}^\prime,\bm{q}_\pi) P_\ell (z) dz\nonumber\\
&=& \frac{1}{2}\int^1_{-1} 
\frac{P_\ell (z) dz}{(p^\prime q_\pi/2\mu) (\beta + z)} \nonumber\\
&=& \frac{(-1)^\ell}{(p^\prime q_\pi/2\mu)} Q_\ell (\beta) \ ,
\end{eqnarray}
with 
\begin{eqnarray}
\beta\equiv \frac{m_\Delta - m_N - p_o^2/m_N + (p^{\prime 2}
 +q^2_\pi/4)/2\mu}
{p^\prime q_\pi/2\mu} \ ,
\end{eqnarray}
and $z\equiv \hat{\bm{p}}^\prime\cdot\hat{\bm{q}}_\pi$.
The function $Q_\ell(\beta)$ is the Legendre function of the second kind,
and is given by
\begin{eqnarray}
Q_\ell (\beta)
&=& \frac{1}{2}\int^1_{-1} 
\frac{P_\ell (z) dz}{ \beta - z} \ .
\end{eqnarray}
I write \Eq{fourier_pi-d} using the expanded form \Eq{exp_deno}.
When I retain only the first term ($\ell=0$) of the expansion, I
obtain
\begin{eqnarray}
\eqn{op_q0}
&&\int \frac{dk^3}{(2\pi)^3} e^{-i\bm{k}\cdot\bm{r}}
 \frac{\bm{k}\; \bm{\sigma}_2\cdot\bm{k}}{m_\pi^{\prime 2} + k^2}\ 
\frac{2\mu}{p^\prime q_\pi} Q_0(\beta) f(k) \nonumber\\
&=&
\left(
\frac{1}{3}\bm{\sigma}_2 F_0 (r) + \frac{\sqrt{8\pi}}{3}
\left[Y_2(\hat{\bm{r}})\otimes\bm{\sigma}_2
\right]_{(1)} F_2(r)
\right)
\frac{2\mu}{p^\prime q_\pi} Q_0(\beta) \ ,
\end{eqnarray}
with
\begin{eqnarray}
F_0 (r) &=& \frac{1}{2\pi^2}\int_0^\infty dk \frac{k^4}{k^2 + m_\pi^{\prime 2}}
j_0 (kr) f(k) \ , \nonumber \\
F_2 (r) &=& \frac{1}{2\pi^2}\int_0^\infty dk \frac{k^4}{k^2 + m_\pi^{\prime 2}}
j_2 (kr) f(k) \ .
\end{eqnarray}
I take a matrix element of \Eq{fourier_pi-d} after setting $q_\pi = 0$,
and compared it with the matrix element of \Eq{op_q0} in which 
$q_\pi (\ne 0)$ is fixed by the kinematics.
In the kinematical region of interest,
I found a small correction ($\sim$ 1.5\%). 
The use of \Eq{op_q0} may be regarded as an inclusion of the lowest
order correction [${\cal O} (q_\pi^2)$] from finite $q_\pi$,
even though there is still another ${\cal O} (q_\pi^2)$ correction.
I do not consider the higher order ($\ell\ge 1$) terms
in \Eq{exp_deno} to be influential on our result because:
the ${\cal O} (q_\pi^2)$ correction from \Eq{op_q0} is small;
the expansion in \Eq{exp_deno} may be regarded as an expansion in terms
of $z/\beta$, and $z/\beta \ll 1$ in most of the kinematical region of
interest.
I use \Eq{op_q0} in our calculation. 
\end{appendix}

\bibliographystyle{unsrt}

\end{document}